\documentclass[a4paper, 11pt]{article}

\usepackage{blindtext}
\usepackage{setspace}
\usepackage{authblk}
\usepackage[caption=false]{subfig}
\usepackage[]{apacite}
\usepackage{rotating}
\usepackage{placeins}
\usepackage[left=3cm, right=3cm, top=3cm, bottom=3cm]{geometry}
\usepackage{amsthm}
\usepackage{amssymb}
\usepackage{amsmath}
\usepackage[utf8]{inputenc}
\usepackage{epstopdf}
\usepackage{amsfonts}

\theoremstyle{plain}

\usepackage{float}

\begin{document}


\title{Modeling partitions of individuals}

\renewcommand\Authfont{\scshape\large}
\renewcommand\Affilfont{\itshape\small}


\author[1]{Marion Hoffman\thanks{Contact: marion.hoffman@gess.ethz.ch}}
\author[2,4]{Per Block} 
\author[3,4]{\\Tom A.B. Snijders}

\affil[1]{Chair of Social Networks, ETH Z\"{u}rich, Weinbergstrasse 109, 8006 Z\"{u}rich, Switzerland}
\affil[2]{Department of Sociology and Leverhulme Centre for Demographic Science, University of Oxford, Oxford OX1 1JD, United Kingdom}
\affil[3]{University of Groningen, Groningen, Netherlands}
\affil[4]{Nuffield College, University of Oxford, Oxford OX1 1NF, United Kingdom}

\renewcommand\Authands{ and }

\date{}

\maketitle

\begin{abstract}

Despite the central role of self-assembled groups in animal and human societies, statistical tools to explain their composition are limited. We introduce a statistical framework for cross-sectional observations of groups with exclusive membership to illuminate the social and organizational mechanisms that bring people together. Drawing from stochastic models for networks and partitions, the proposed framework introduces an exponential family of distributions for partitions. We derive its main mathematical properties and suggest strategies to specify and estimate such models. A case study on hackathon events applies the developed framework to the study of mechanisms underlying the formation of self-assembled project teams.

\end{abstract}


\section{Introduction}
\label{sec:Introduction}

\subsection{The study of self-assembled groups}
\label{subsec:The study of self-assembled groups}

In gregarious species, individuals have a tendency to come together in groups. This is especially pertinent in humans. Often, the composition of these groups emerge from voluntary decisions of members, thus, crystallizing socializing preferences in social groups, or goal-oriented behaviors in the case of task-oriented groups. 
In some cases, membership to a group is exclusive, in the sense that every individual can only be member of one group. 
This exclusivity might result from physical and temporal constraints---e.g. when group boundaries are defined by physical gathering---or structural rules---e.g., group overlap is often forbidden when groups compete for some goal. 
In such cases of self-assembled and exclusive groups, the decision to group with certain individuals rather than others can depend on important social mechanisms that structure the organisation of a community. The present paper introduces a statistical framework to model and explain observations of self-assembled exclusive groups, with a view to better understand the mechanisms underlying their formation.

Examples of self-assembled exclusive groups are numerous, ranging from mammal herds in the wild to player squads in online games. Numerous situations require individuals to organize themselves into such groups, in order for them to execute an action or acquire a resource. In the animal kingdom, many species gather into flocks, herds, or schools for traveling purposes \shortcite{Okubo1986-md, Reynolds1987-ug} and predators assemble packs for hunting \shortcite{Creel1995-ij,Gittleman1989-ar}. 
In the human world, children groups gather in the schoolyard to engage in common activities \shortcite{Moody2001-rp}, sport clubs  emerge to provide opportunities for shared free-time activities \shortcite{Putnam2000-xh,lazarsfeld1954friendship}, and project teams assemble spontaneously to tackle organizational tasks \shortcite{guimera2005team,zhu2013motivations}. In this paper we use the example of human groups and in particular project teams in the empirical illustration.

The existence and composition of social groups have a crucial role in determining societal outcomes. Seminal sociological works recognize that by coming together in groups, individuals influence each other's cognition, affective structures, and individual outcomes \shortcite{Parsons1949-et,homans1950human,lewin1936principles}. 
Various theories develop concepts for group settings, such as social circles \shortcite{Simmel1949-tz}, social foci \shortcite{Feld1982-bj}, social settings \shortcite{Pattison2002-xb}, or social situations \shortcite{Block2018-wk}. 
Groups lay ground for the development of social ties \shortcite{Fischer1982-ut, Moody2001-rp, lazarsfeld1954friendship,Simmel1949-tz} and provide the context for exchange relations \shortcite{granovetter1985economic}, where the ability of one group member to acquire resources and social support will depend on what the other members can provide. 
Additionally, the set of attributes present in a group as well the relations between its members might have an impact on some essential group outcomes. 
At a broader level, the formation of groups in a community can indicate and impact how different parts of the community relate to each other and segregate \shortcite{Allport1954-pv}. Investigating group formation is all the more important in instances where such outcomes are crucial to the functioning of individuals and communities. The study of these interdependent group processes calls for the development of mathematical tools tailored for this level of social unit, as argued by Lindenberg \citeyear{Lindenberg1997-gc}.

The main aim of the model we propose is to uncover which mechanisms guide the composition of self-assembled groups in a given setting, and to assess their relative importance.
Such mechanisms can fall in the categories of biological imperatives, social preferences, and exogenous constraints. 
Adding to the variety of their origin, the mechanisms underlying group formation can also be situated at different levels.
\begin{enumerate}
\setlength\itemsep{0mm}
\item For any group member, the characteristics of the other members reflect their individual attraction towards others exhibiting some particular attribute. 
\item Group composition can also reflect dyadic preferences, such as the preference of individuals connected through a relationship (e.g., kinship or friendship) to belong to the same groups. 
\item Finally, group-level mechanisms, such as the optimization of a certain combination of attributes, can guide group formation. 
\end{enumerate}
In the example of project teams, individuals might seek (1) teams with individuals who have similarities that promote mutual understanding and common expectations, (2) colleagues with whom they have already collaborated, or (3) teams with an efficient distribution of competences \shortcite{skvoretz2016red}. On top of these formation mechanisms, some contexts might constrain group compositions or sizes -- for example, a maximal group size might be imposed. The proposed model aims to shed light on the role of these diverse factors in group formation processes while taking such constraints into account.

\subsection{Previous approaches}
\label{subsec:Previous approaches}

\subsubsection{Network approaches}
A common approach to represent group membership is to define a two-mode, or bipartite, network in which nodes on one level (i.e., individuals) are connected to a second level of nodes (i.e., groups). We review here the use and limitations of models for such representations.

First, permutation test techniques and models such as the Quadratic Assignment Procedure (QAP) proposed by \shortcite{Krackhardt1988-ex} can be used to investigate whether some combinations of attributes within groups are more likely than others in a bipartite network. However, these models condition on group structure (i.e., the distribution of group structures) and cannot be used to investigate the factors explaining the distribution of group sizes. Moreover, interpreting the effect of covariates conditional of group structure can be problematic when these covariates are potentially responsible for the group structure itself (e.g., if an attribute explains both the size of groups and homophily within the groups).

Alternatively, the Exponential Random Graph Model (ERGM) leverages the capabilities of exponential family models \shortcite{sundberg2019statistical} and make use of techniques from spatial statistics \shortcite{besag1974spatial} and graphical modeling \shortcite{lauritzen1996graphical} to capture more complex dependencies between the membership ties \shortcite{lusher2013exponential,schweinberger2020exponential}.  
The ERGM can be used to model both attribute and structural dependencies, such as the propensity of individuals to join groups in case they already share other groups with their members. 
Theoretically, it is possible to restrict the support of an ERGM to bipartite networks with individuals' degrees fixed to one \shortcite{Morris2008-pi}, thus allowing to model exclusivity in group membership, and create some structural effects to capture group sizes. 
The main problem remaining is that the number and characteristics of second mode nodes should be predetermined and cannot be modelled themselves \shortcite{wang2009exponential}. A way to circumvent this could be to set an initial number of second mode nodes equal to the maximum number of groups, i.e., the number or actors. In that case, it is straightforward to see that different partitions would be represented by different numbers of equivalent networks. For example, for three individuals A, B, and C, the partition with three isolated individuals would be equivalent to six bipartite networks, while the partition with A, B, and C in the same group would only be equivalent to three networks.
Consequently, interpreting the structural parameters of such a constrained ERGM would be problematic. All in all, using constrained ERGMs would bring few insights into the mechanisms underlying the number and size of the self-emergent groups.

A variation of the network logic that integrates the constraint of exclusive group membership was defined in the general location system (GLS) model of \citeauthor{butts20079} \citeyear{butts20079}. This model is tailored to observations where individuals can be assigned to only one group (or location) at a time, similarly to how individuals set themselves into occupations or geographical residences. In the same vein as the ERGM, the GLS framework builds upon the exponential family formalism but requires to know the number and characteristics of the available groups in advance.

So far, only approaches designed for dynamic changes in group compositions over time could circumvent the issue of having to pre-define the second mode nodes, by artificially creating and deleting the second mode nodes \shortcite{Hoffman_2020}, but such procedures remain ill-suited for cross-sectional observations. 

\subsubsection{Partition approaches}
A way to circumvent the difficulty of not having predefined groups is to represent groups as a partition of the set of individuals, with a partition being a division of the individuals into non-overlapping groups. Popular partition distributions are the uniform Dirichlet-multinomial partitions defined for partitions with a maximum possible number of groups \shortcite{mccullagh2011random,kingman1978random} and Poisson-Dirichlet distributions \shortcite{pitman1997two}. Such families of distributions still assume a predefined number of available classes, although the possible number of groups now sits between one and a maximal value. The extension of these models when the maximum value becomes infinite is known as the Ewens distribution \shortcite{ewens1972sampling,mccullagh2011random}. 
The Ewens distribution was first applied to the problem of allele sampling in genetics \shortcite{ewens1972sampling}, but its use, as well as the use of the related Dirichlet distributions, has spread into the fields of biodiversity \shortcite{Hubbell2001-sk}, Bayesian statistics \shortcite{Ferguson1973-sl,Antoniak1974-jk} and many other fields of mathematics \shortcite{crane2016ubiquitous}. Interestingly, the Ewens specification also defines an exponential family \shortcite{crane2016ubiquitous}. 
One limitation of these models is that they cannot incorporate attribute and structural dependencies between group memberships in the same way ERGMs and GLS models do. This is connected to their main applications to sampling problems.

In this paper, we incorporate insights from the network and partition modeling literature into a novel statistical framework suited for observations of self-assembled and exclusive groups. This framework represents groups of individuals as a partition of a set of individuals and builds upon the literature on exponential families for networks to capture non-trivial dependencies between the groups composing the partition. The model allows for the size and composition of groups to be the result of individual, relational, and group-level processes and offers the possibility to draw inference on the processes driving the formation of groups in a certain context. Sections 2 to 4 describe the definitions, mathematical formulation, and interpretation of the model. Sections 5 and 6 cover the computation and estimation of the model parameters. Section 7 presents an application to the study of self-assembled teams during hackathon competitions.

\section{Definitions}
\label{sec:Model definition}

\subsection{Notation}
\label{subsec:Notation}

Consider a set of $n$ actors $\mathcal{A}$. A \textit{partition} $P$ over $\mathcal{A}$ represents a division of these actors into non-overlapping subsets. Formally, $P$ is a set of \textit{groups} or \textit{blocks}, denoted $G$, that satisfies the conditions:
\begin{gather}
    \bigcup_{G \in P} = \mathcal{A}, \notag \\[1pt]
    \forall (G,G') \in P^{2}, G \neq G': \; G \cap G' = \emptyset, \notag \\[4pt]
    \forall G \in P: \; G \neq \emptyset. \notag
\end{gather}
For convenience, we define the function $g_{P}: [\![ 1,n ]\!] \mapsto P$ returning the group of a given node:
\begin{equation}
    g_{P}(i) = G \; | \; i \in G.
\notag
\end{equation}
We can also transform the partition representation into the binary $n \times n$ matrix
\begin{equation}
 X = \big[ x_{i,j} \big]_{i,j \in \mathcal{A}} 
 \textnormal{ where } x_{i,j}=1 \Leftrightarrow g_{P}(i) = g_{P}(j). \notag
\end{equation}
Figure 1 illustrates different possible representations of a partition in comparison to the ones used in the case of networks.

\begin{figure}[t]
    \centering
    \vspace{10pt}
    \includegraphics[scale=0.48]{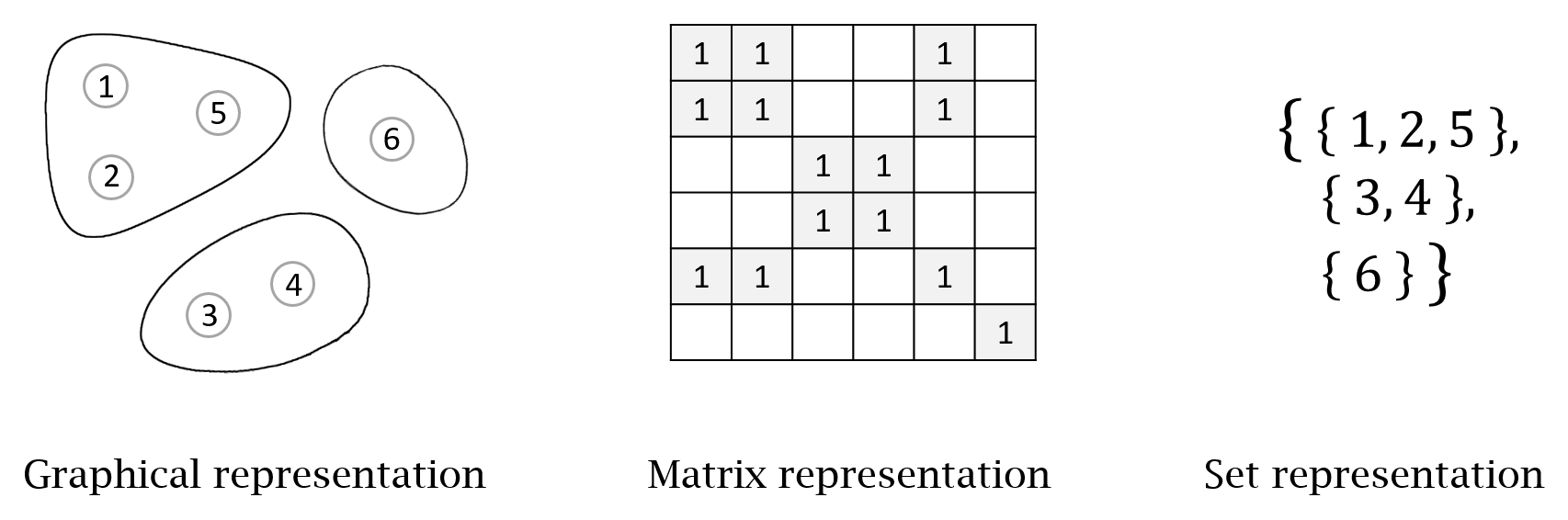}
    \caption{Possible representations of a partition over the nodeset $\{1,2,3,4,5,6\}.$}
    \label{fig:illustration}
    \vspace{5pt}
\end{figure}


In the following sections, we use the notation $\#P$ for the number of groups in a partition $P$, and $\#G$ to define the size of a given group $G$. Furthermore, we use the letter $P$ when referring to a random partition, and $p$ for the realization of a partition. To avoid any confusion, probabilities are written with the symbol $\textnormal{Pr}$.

We further use the letters $a$ for individual covariates (e.g. gender, age) and $Z$ for dyadic covariates (e.g., friendship ties).

\subsection{Definition of the partition set}
\label{subsec:Definition of the partition set}

The power set of all partitions over the set $\mathcal{A}$ is referred as $\mathcal{P}(\mathcal{A})$ (or $\mathcal{P}$ when the nodeset is not ambiguous). The size of $\mathcal{P}$ is given by the Bell number $B_n$ \shortcite{bell1934exponential,pitman1997some} and can be calculated iteratively by:
\begin{equation}
    B_{0} = 1 \textnormal{ and }
    B_{n+1} = \sum_{i=0}^{n}{\binom{n}{i}B_{i}}. 
\label{equ:bell_number}
\end{equation}

In certain contexts, some partitions of the actor set might not be realistic or allowed, in which case one might only consider a subset $\mathcal{P}'$ of the whole partition space $\mathcal{P}$.
Most prominently, certain group sizes might not be allowed. When considering subsets $\mathcal{P}'$ that only contain groups of sizes higher or equal to minimal value $s_{min}$ and lower or equal to a value $s_{max}$, a number of calculations can be extended. The number of partitions belonging to this subset can be calculated similarly to Equation \eqref{equ:bell_number} with a sequence $B'_{n}$ (details can be found in Appendix A). After defining the values $i_{min} = \max(0, n+1-s_{max})$ and $i_{max}=\min(n, n+1-s_{min} )$, $B'_{n}$ is defined by:
\begin{gather}
    B'_{n} = 0 \textnormal{ for } 0 < n < \sigma_{min}, \notag \\
    B'_{0} = B'_{\sigma_{min}} = 1, \notag \\
    B'_{n+1} = \sum_{i=i_{min}}^{i_{max}} {\binom{n}{i} B'_{i}} \textnormal{ for } n \geqslant \sigma_{min}. 
    \notag
\label{equ:bell_number_2}
\end{gather}

\subsection{Relations between partitions}
\label{subsec:Relations between partitions}

For the purpose of parameter estimation and interpretation, we define three symmetric binary relations between the elements of $\mathcal{P}$, called the \textit{merge/split}, \textit{permute}, and \textit{transfer} relations (see an illustration of these relations in Figure 2). 

The \textit{merge/split} relation $\mathcal{R}^{\textit{merge}}$ is the set of all unordered pairs of partitions for which one partition of the pair is obtained by merging two distinct groups in the other partition. Since these are unordered pairs, in the reverse direction this definition includes splitting one group in one partition into two groups in the other. Formally, we define $P_{-G,G'}=P \setminus \{G,G'\}$ the partition $P$ with two groups $G$ and $G'$ removed. The relation can be written as:
\begin{equation}
    \mathcal{R}^{\textit{merge}} = \big\{ \{P,P'\} \subseteq \mathcal{P} \; | \; \exists G,G' \in P \textnormal{: } P' = P_{-G,G'}  \cup \{G \cup G'\} \big\}.
\notag
\end{equation}

The \textit{\textit{permute}} relation $\mathcal{R}^{\textit{permute}}$ links partitions in which two nodes in two different groups are exchanged, while the other nodes grouping remains the same. For $i$ and $i'$ two nodes respectively belonging to two distinct groups $G$ and $G'$, we note $G_{i \leftrightarrow i'}$ and $G_{i' \leftrightarrow i}'$ the groups in which the nodes $i$ and $i'$ have been exchanged. Under the same notation the relation defines the following unordered pairs: 
\begin{equation}
    \mathcal{R}^{\textit{permute}} = \big\{ \{P,P'\} \subseteq \mathcal{P} \; | \; \exists G,G' \in P \textnormal{, } i \in G \textnormal{, } i' \in G' \textnormal{: } P' = P_{-G,G'}  \cup \{G_{i \leftrightarrow i'}, G_{i' \leftrightarrow i}'\} \big\}.
\notag
\end{equation}

Finally, the \textit{\textit{transfer}} relation $\mathcal{R}^{\textit{transfer}}$ contains the unordered pairs of partitions $\{P,P'\}$ for which partition $P$ and $P'$ are identical, with the exception of one node that belongs to a different group in $P$ and $P'$ (we can say that this node is being transferred from one group to another). Importantly, this node may be an isolate in one of the two partitions. Similarly, for a node $i$ belonging to the original nodeset $\mathcal{A}$, we denote $P_{-i}$ the projection of the partition on the set $\mathcal{A} \backslash \{i\}$. The relation is then defined by: 
\begin{equation}
    \mathcal{R}^{\textit{transfer}} = \big\{ \{P,P'\} \subseteq \mathcal{P} \: | \: \exists i \in \mathcal{A} \textnormal{: } P'_{-i} = P_{-i}  \textnormal{ and } P' \neq P \big\}.
\notag
\end{equation}

\begin{figure}[t]
\centering
	\includegraphics[scale=0.5]{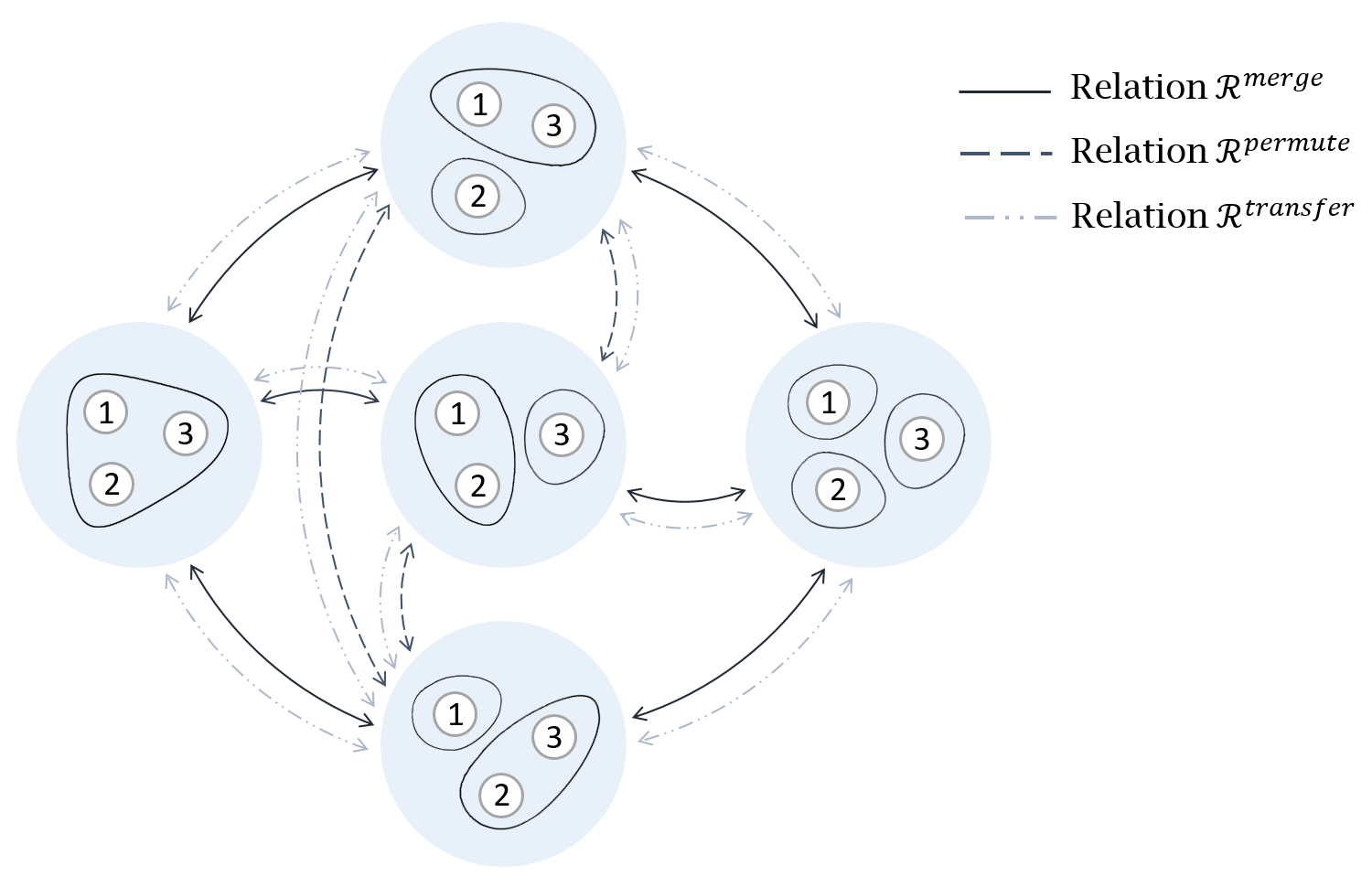}
    \caption{Illustration of the \textit{merge/split}, \textit{permute}, and \textit{transfer} relations for the full set of partitions over three nodes. All relations are binary and symmetric.}
    \label{fig:relation_illustration}
\end{figure}

\subsection{Definition of the probability distribution}
\label{subsec:Definition of the probability distribution}

Our aim is to define a parametric set of probability distributions over $\mathcal{P}$ for a given set of actors. 
The parameters of this distribution should be associated to statistics relevant for the hypotheses under consideration on the processes resulting in the observed partition. As outlined earlier, such hypotheses can be associated with the structure of the partition (i.e. the number of groups and their sizes) or the distribution of actors' attributes within the groups. 

The class of exponential distributions allows such parametrization in a straightforward way \shortcite{sundberg2019statistical}. We propose here an exponential family with support the set $\mathcal{P}$ (or a subset $\mathcal{P}'$). This family is defined for an identity base measure, a vector of natural sufficient statistics $s(P)= \big( s_{k}(P) \big)_{k \in K}$, and a canonical parameter vector $\alpha= \big( \alpha_{k} \big)_{k \in K}$. It is expressed by:
\begin{equation}
    \textnormal{Pr}_{\alpha}(P = p)
    = \frac{\exp{ \Big( \sum_{k}{\alpha_{k} s_{k}(p)} \Big) } }
    {\kappa_{\mathcal{P}}(\alpha)}.
\label{equ:ERPM_jointform}
\end{equation}

where the normalizing constant $\kappa_{\mathcal{P}}(\alpha)$ is defined by:
\begin{equation}
    \kappa_{\mathcal{P}}(\alpha) = \sum_{\widetilde{P} \in \mathcal{P}}{\exp{\Big(\sum_{k}{\alpha_{k} s_{k}(\widetilde{P})}\Big)} }.
\label{equ:ERPM_kappa}
\end{equation}

This formulation mirrors the definition of an ERGM when considering a partition instead of a graph distribution (see \shortcite{lusher2013exponential} or \shortcite{robins2007introduction} for more details on ERGMs). 

Some special cases of this exponential family are related to well-known distributions. Naturally, the model defined without any sufficient statistic generates the uniform distribution over the partition set $\mathcal{P}$. The Ewens distribution \shortcite{ewens1972sampling,mccullagh2011random} is defined for a positive parameter $\lambda$ as follows:
\begin{equation}
    \textnormal{Pr}_{\lambda}(P=p) = 
    \frac{\Gamma(\lambda-1) \; \lambda^{\#p} \; \prod_{G \in p}{(\#G-1)!}}{\Gamma(n+\lambda-1)}
\label{equ:Ewens}
\end{equation}
with $\Gamma$ being the Gamma function. 
As shown in Appendix B, this definition is equivalent to the following formulation of \eqref{equ:ERPM_jointform} with the parameter vector $\alpha = (\log(\lambda),1)$:
\begin{equation}
    \textnormal{Pr}_{\alpha}(P = p) = 
    \frac{\exp \Big( \alpha_{1} \: \#p + 
    \alpha_{2} \: \sum_{G \in p}{\log \big( (\#G-1)! \big) } \Big)}
    {\kappa_{\mathcal{P}}(\alpha)}.
\label{equ:Ewens2}
\end{equation}




\section{Model Specification}
\label{sec:Model Specification}

\subsection{Sufficient statistics}
\label{subsec:Sufficient statistics}

Graphical modeling with dependence graphs is a useful technique for specifying exponential family distributions \shortcite{lauritzen1996graphical}. In the network literature, this technique was introduced for Markov graphs by \citeauthor{frank1986markov} \citeyear{frank1986markov} and later developed for ERGMs \shortcite{wasserman1996logit,robins2005interdependencies}. Dependence graphs then capture the dependence structure of the tie variables and this structure can inform the choice of relevant sufficient statistics, by virtue of the Hammersley-Clifford theorem \shortcite{hammersley1971markov,besag1974spatial}. 
However, graphical modeling is ill-suited to study partition models since the dependence graph of group variables with the non-overlapping constraint is not straightforward. Instead, we take inspiration from the statistics and the independence assumptions used in other related statistical models, in particular partition models (i.e., Ewens and Dirichlet partitions) or Dirichlet models. 
Extending the statistics used in the Ewens formula, we show that statistics defined as sums of group attributes can model a wide range of partition properties. The independence properties of count statistics are described in Section \ref{subsec:Independence properties of the distribution}.

\subsubsection{Structural statistics}
\label{subsubsec:Structural statistics}

Structural statistics aim to accurately model the observed group sizes and their dispersion in a given partition. To understand which statistics can be used, we calculate the expected distribution of group sizes in random partitions of $10$ nodes for different statistics related to group sizes. Having $n=10$ allows us to enumerate the number of partitions with specific group statistics and directly calculate their probabilities (for more details of these probabilities, see Section \ref{sec:Computation of the normalizing constant of the distribution}). Increasing the number of nodes does not affect the behavior of the presented statistics.

\begin{figure}
	\includegraphics[width=1\textwidth]{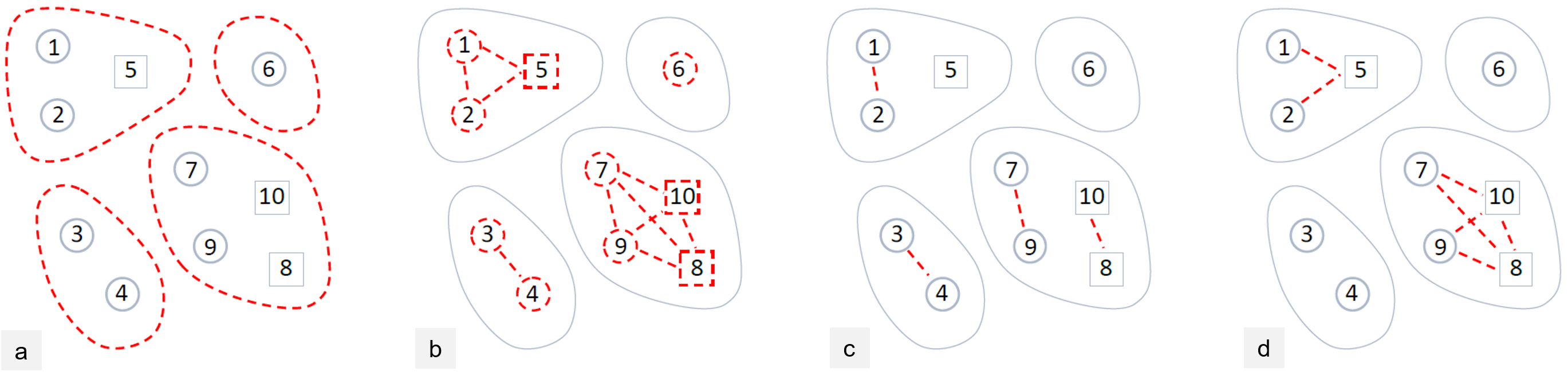}
	\caption{Illustration of the calculation of introduced statistics based on counts for a given partition $p=\big\{ \{1,2,5\},\{3,4\},\{6\},\{7,8,9,10\} \big\}$ with a binary covariate (actor's shape); the red dashed elements are to be counted to get the statistics value for (a) number of groups $s_{1}(p)=4$; (b) squared group sizes, i.e.\ each unordered dyad must be counted twice plus the number of nodes $s_{3}(p)=30$; (c) number of dyads within groups that are identical on shape $s_{\text{dyadic homophily}}(p)=4$; and (d) number of ordered dyads within groups that include one square $s_{\text{dyadic sociability}}(p)=8$.}
	\label{fig:effect_stat_illustration}
\end{figure}

\begin{figure}
    \includegraphics[scale=0.4]{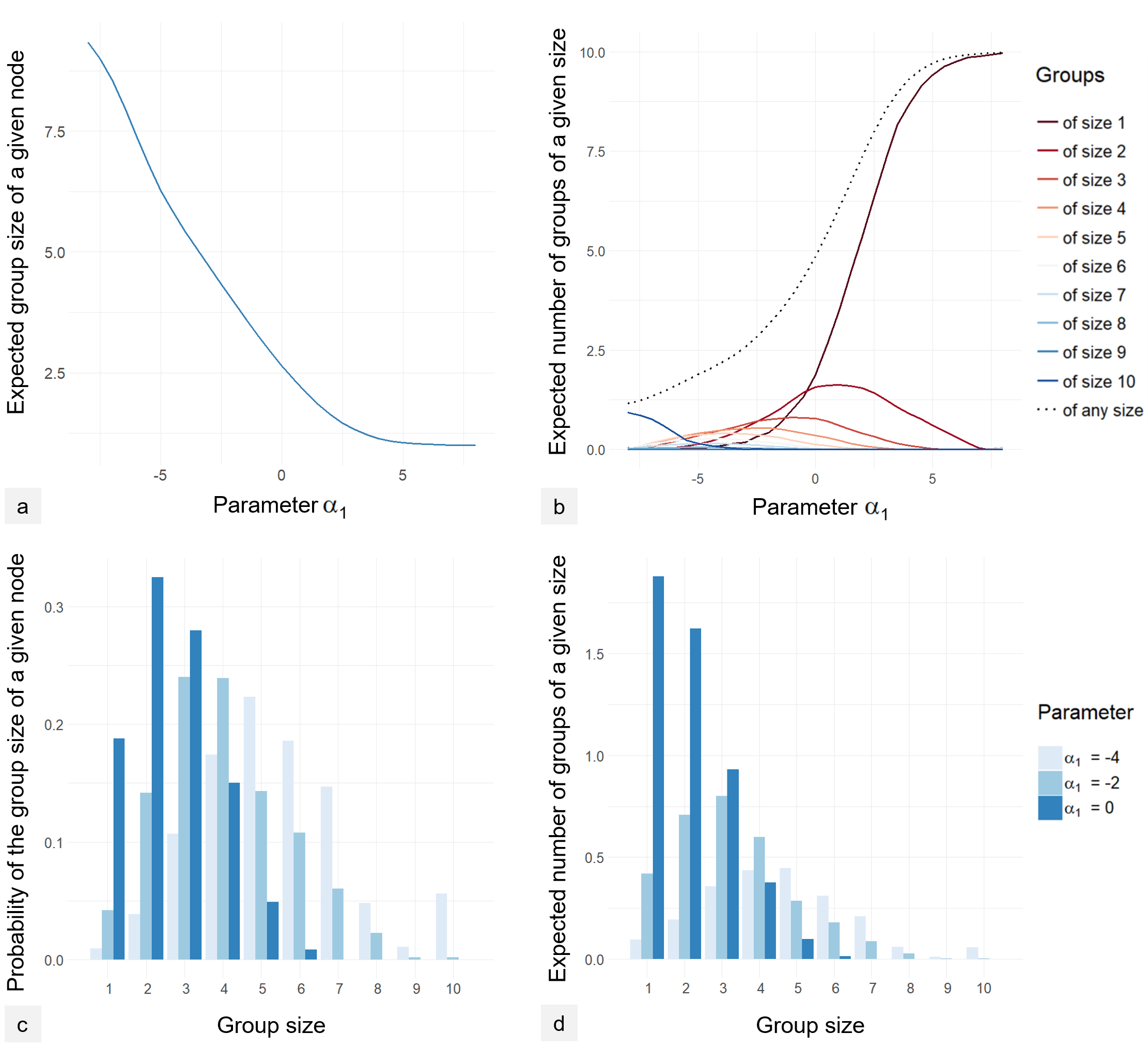}
    \caption{Distribution of group sizes in a random partition defined by a model with $10$ nodes and one sufficient statistic $s_{1}(P)=\#P$, as a function of the parameter $\alpha_{1}$. (a) Expected group size of a given node; (b) expected number of groups of a given size; (c) probability function for the size of a given node for three values of $\alpha_{1}$; and (d) expected distribution of group sizes for three values of $\alpha_{1}$.}
        \label{fig:simulations_numgroups}
\end{figure}

The first relevant statistic to model group sizes is the number of groups (i.e., the cardinality of the partition, see Figure \ref{fig:effect_stat_illustration}a)
\begin{equation}
    s_{1}(P) = \#P,
\notag
\end{equation}
as it is the basis of the Ewens formula (see Equation \eqref{equ:Ewens2}). Figure \ref{fig:simulations_numgroups}a shows that low values of $\alpha_{1}$ favor partitions with large groups of $10$, $9$, or $8$ nodes, while high values favor many small groups of $1$, $2$ or $3$ nodes. 
Figure \ref{fig:simulations_numgroups}b shows that, as $\alpha_{1}$ increases, the expected number of groups increases, and so does the expected number of singleton groups. The expected number of groups of size 10, i.e., trivial one-group partitions, decreases, while the expected prevalence of the intermediate group sizes 2—9 is unimodal, and assume their maxima for values of $\alpha_{1}$ that decrease with group size.
Figure \ref{fig:simulations_numgroups}c further shows that the probability for a random node to belong to large groups decreases when $\alpha_{1}$ increases. 
Finally, Figure \ref{fig:simulations_numgroups}d shows that the distribution of group sizes stochastically decreases with $\alpha_{1}$. 
We conclude that the number of groups is a simple and efficient way to model the central tendency of group sizes in a partition.

\begin{figure}
    \includegraphics[scale=0.41]{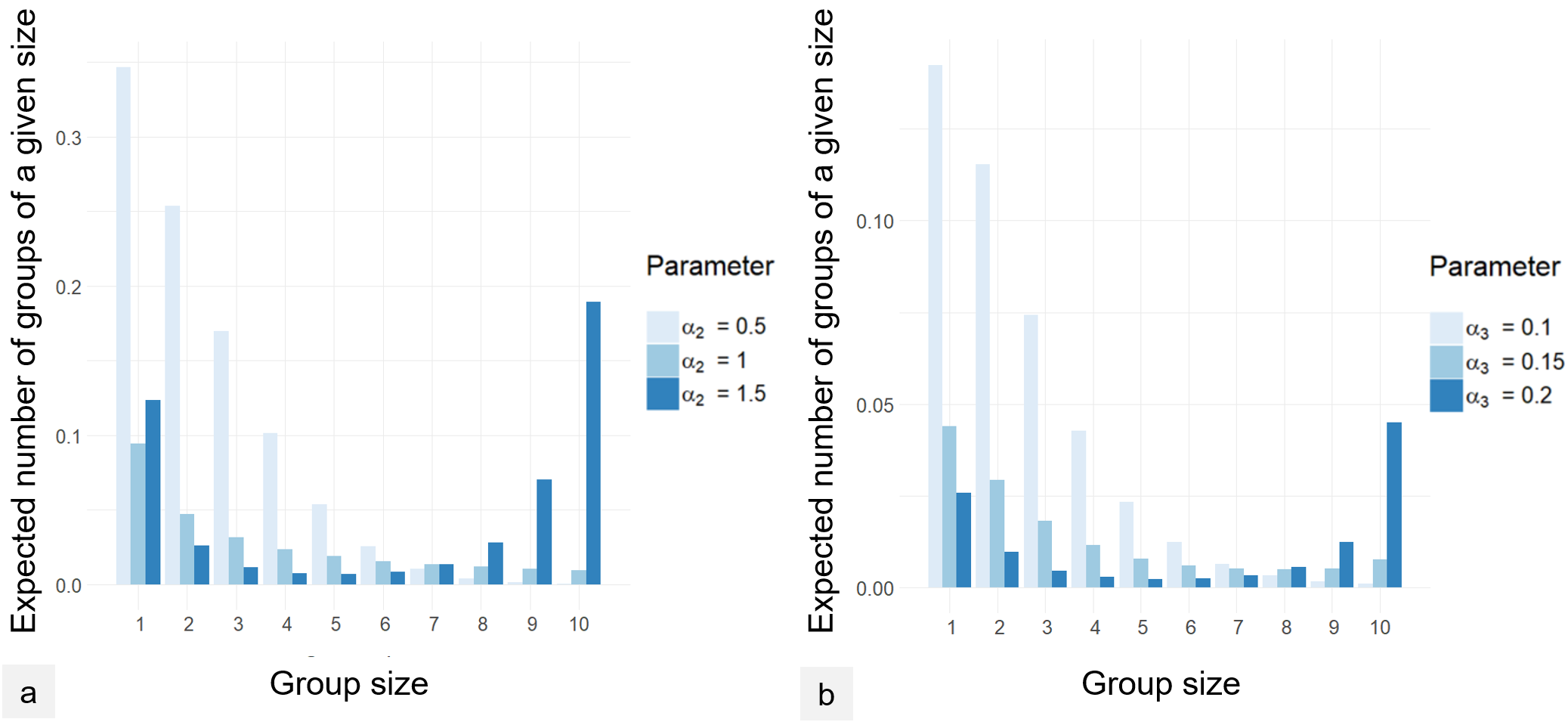}
    \caption{Distribution of group sizes for a random partition defined by a model with $10$ nodes and two sufficient statistics for three values of the second parameter (see the text for the determination of the first parameter). (a) $s_{1}(P)=\#P$, $s_{2}(P) = \sum_{G \in P}{\log \big( (\#G-1)! \big)}$; and (b) $s_{1}(P)=\#P$, $s_{3}(P) = \sum_{G \in P}{\#G^2}$.}
    \label{fig:simulations_dispersionnumgroups}
\end{figure}

Another important feature of the group size distribution is its dispersion or skewness. We first use the statistic in Definition \eqref{equ:Ewens2} of the Ewens distribution: 
\begin{equation}
    s_{2}(P) = \sum_{G \in P}{\log \big( (\#G-1)! \big)}.
\notag
\end{equation}
Since the Ewens model can reproduce a "richer-get-richer" effect on group sizes \shortcite{mccullagh2011random}, meaning that sizes become exponentially distributed, we can expect it is the result of this term being included in the model. 
We calculate the size distribution for partitions over $10$ nodes for a model containing the two statistics $s_{1}$ and $s_{2}$ by varying the parameter $\alpha_{2}$. 
To fix the first statistic, we determine the value $\alpha_{1}$ that maintains the expected value of $s_{1}$ equal to $4$ for each pre-determined $\alpha_{2}$.
This means we explore the distribution of expected group sizes for a constant expected number of groups.
Figure \ref{fig:simulations_dispersionnumgroups}a shows the expected distribution.
The dispersion of sizes increases with the parameter value for the statistic $s_{2}$. 

Another intuitive statistic for modeling the skewness of the size distribution is the sum of squared sizes:
\begin{equation}
    s_{3}(P) = \sum_{G \in P}{\#G^2}.
\notag
\end{equation}
It is equal to the sum of the elements of the matrix representation $X$ of the partition (see Figure \ref{fig:effect_stat_illustration}b).
The group size distributions obtained for this statistic are shown in Figure \ref{fig:simulations_dispersionnumgroups}b. 
Once again, increasing the value $\alpha_{3}$ can increase the dispersion of sizes in the random partition. Choosing between $s_{2}$ and $s_{3}$ to model size dispersion is then a practical matter of which one represents more accurately the structure of the observed partition.

In case the distribution of group sizes cannot be approximately reproduced by the above parameters, or if a particular group size might be over- or under-represented for exogenous reasons, the number of groups of particular sizes can be added as a sufficient statistics.

At this point, it is important to mention that some estimation issues coined as \textit{degeneracy} or \textit{near-degeneracy} by the ERGM literature \shortcite{handcock2003assessing,snijders2006new,robins2007recent,lusher2013exponential} might ensue from the use of certain statistics combinations in this model. This is the case for the previous models defined for $S=\big( s_1,s_2 \big)$ and $S=\big( s_1,s_3 \big)$.
As a result, some estimated models will correspond to unrealistic distributions that concentrate their probability mass on a few extreme partitions such as the one with only one group and the one only containing singletons rather than accurately reflecting the observed statistics.
In most cases, a degenerate model will point to some misspecification, and it might prove useful to have a different operationalization of size dispersion. For example, one might use a weighted sum over all group sizes of the number of cliques of a given size. Weights could be defined as decreasing in a similar way as the "geometrically weighted edgewise shared partners" (gwesp) effect proposed in \citeauthor{snijders2006new} \citeyear{snijders2006new} and \citeauthor{hunter2007curved} \citeyear{hunter2007curved}. Most observations on degeneracy made in the case of ERGMs can be extended to the model presented here. 

\subsubsection{Statistics for covariates}
\label{subsubsec:Statistics for covariates}

The influence of individual covariates on the formation of relational ties has been widely investigated in social networks, starting with the fundamental idea of \textit{homophily} \shortcite{mcpherson2001birds,rivera2010dynamics} stating that similar individuals are more likely to be connected. 
In the case of a dyad, homophily can be operationalized as a dyadic variable indicating whether the actors have the same (or similar) attributes. 
Including this mechanism in the current model requires to extend the concept of homophily to the group level.

The most direct extension of dyadic homophily is to consider a group as homophilous if the similarity within all dyads of actors within the group is high. For example, hackathon participants might be more likely to form teams in which everyone is about the same age. 
If we define a dyadic similarity index $\text{sim}_{i,j}$ for two actors $i$ and $j$, we can operationalize such type of homophily by counting the sum of similarity indexes among dyads in all groups:
\begin{equation}
    s_{\text{dyadic homophily}}(P) = \sum_{G \in P}{ \sum_{i,j \in G}{\text{sim}_{i,j}} }.
\notag
\end{equation}

For a binary attribute, the similarity index can take the value of $1$ when two actors have the same attribute, and $0$ otherwise. This amounts to counting the number of ties between identical individuals in the network representation of the partition, as illustrated in Figure \ref{fig:effect_stat_illustration}c. For a categorical and ordered attribute, or a continuous attribute, we can simply use the absolute difference between actors' attributes (in that case, a positive parameter for this effect shows heterophily, and a negative one homophily). 

Alternatively, homophily can operate at the group level. That would mean that the similarity of dyads does not matter as much as the distribution of the attributes within groups. For example, hackathon participants might form teams that are strictly non-mixed, in terms of language spoken or gender. In that case, a similarity index $\text{sim}_G$ is defined for any group $G$, and the relevant statistic becomes:
\begin{equation}
    s_{\text{group homophily}}(P) = \sum_{G \in P}{ \text{sim}_G}.
\notag
\end{equation}

For a binary attribute, the similarity index of a group can be a simple indicator that equals $1$ when all group members have the same attribute. One can also use the range of the attribute values in the group in the case of a continuous attribute (as an indication of the concentration of values in a certain interval) or the number of different values of the attribute in the group in the case of a categorical variable. Positive parameters for the last two implementations would again show heterophily, similarly to using absolute differences above. Alternatively, the within-group variance of the attribute or, the product $p(1-p)$, where $p$ is the proportion of a binary attribute, can also be good choices for $\text{sim}_G$. A negative parameter would again indicate a tendency for low variance within groups and therefore some kind of homophily. 

Many variations on the idea of similarity (or dissimilarity) within group members can be constructed. In the dyadic definition, the count of similar ties could be replaced by the count of individuals who have at least a certain number of similar individuals in their groups, if similarity between group members only matters until a certain threshold. In the group definition, one can consider a statistic counting groups with a certain combination of attributes, if complementarity is thought to be more important than similarity.

In practice, these definitions might lead to include a bias towards certain numbers of groups (in the case of $s_{\text{group homophily}}$) or certain sizes of groups (in the case of $s_{\text{dyadic homophily}}$). It can be important either to control for the number of groups and ties (with the structural statistics $s_1$ and $s_3$) or to normalize these statistics by the number of dyads in each group (e.g., for $s_{\text{dyadic homophily}}$) or the size of the group (e.g., for $s_{\text{group homophily}}$) to increase comparability, especially in empirical cases of heterogenous group sizes.

Finally, the effect of a dyadic covariate $Z$ (e.g., friendship), can simply be added as:
\begin{equation}
    s_{\text{dyadic covariate}}(P) = \sum_{G \in P}{ \sum_{i,j \in G} {Z_{i,j}} }.
\notag
\end{equation}

\subsubsection{Mixed statistics}
\label{subsubsec:Mixed statistics}

Mechanisms related to both covariates and structural features can be included in the current framework. 
This includes a translation of the network concepts of \textit{sociability} or \textit{aspiration} \shortcite{snijders2019beyond}, defined as the tendencies for actors with a high attribute to send or receive more ties, respectively.
For groups, these mechanisms translate into the preference of actors that score high on an attribute to be in larger groups. 
For example, extraverted individuals might be more likely to be found in larger groups, which can be modeled by counting intra-group ties to these individuals with the following sociability statistic:
\begin{equation}
    s_{\text{dyadic sociability}}(P) = \sum_{G \in P}{ \sum_{i \in G}{(\#G-1) \; a_{i}} }.
\notag
\end{equation}
This is illustrated by Figure \ref{fig:effect_stat_illustration}d. 
Alternatively, the sum of individual attributes can be replaced by the average value of the attributes within groups to switch to a group-level definition:
\begin{equation}
    s_{\text{group sociability}}(P) = \sum_{G \in P}{ \#G \; \textnormal{mean} \big[ a_{i} \big]_{i \in G} }.
\notag
\end{equation}

\subsection{Independence properties of the distribution}
\label{subsec:Independence properties of the distribution}

Classical methods of graphical modeling \shortcite{lauritzen1996graphical} are ill-suited for representing dependence assumptions in partition models. However, we can use other concepts to discuss independence properties of the model.

Kingman \citeyear{kingman1978random} established the property of \textit{consistency} for the Ewens sampling formula.
This concept represents that for a given sampled population that can be modeled with an Ewens distribution of parameter $\lambda$, any sub-population of this population also follows an Ewens distribution of the same parameter.
As shown by simple counter-examples in Appendix C, most models defined for the statistics presented above fail to fulfill this condition. The Ewens formula is a special case in that regard, since consistency is a critical property for the study of population samples and much less so in the case of complete observations in our case.

A second relevant concept is \textit{neutrality}, as introduced by Connor and Mosimann \citeyear{connor1969concepts} to study distributions of proportions of a fixed quantity. Such variables are defined as a strictly positive vector $(X_1,X_2,...,X_n)$ with $X_1 + X_2 + ... + X_n = q$ where $q$ is constant. Each variable will never be independent from the others as it can be expressed as a linear combination of the others.
To remedy this, \shortciteauthor{connor1969concepts} introduce the concept of neutrality that defines that the proportion $X_1$, for example, is neutral if it is independent of the vector $\big( X_2 / (q-X_1),...,X_n / (q-X_1) \big)$. This property allows to ignore one or several proportions to study the others. For example, it was shown that neutrality of all proportions characterizes the Dirichlet distribution \shortcite{connor1969concepts, geiger1997characterization}.

Although the concept of neutrality was initially defined for proportion vectors, its extension to partitions can help us understand how the composition of a subset of the partition might affect the rest of the partition. Let $P$ be a random partition over a set $\mathcal{A}$, and $\mathcal{A}'$ a subset of $\mathcal{A}$, with complement set $\mathcal{A}'^{c}$. We further define $\pi$ and $\pi^{c}$ as the respective projections of partitions in $\mathcal{P}(\mathcal{A})$ over $\mathcal{A}'$ and $\mathcal{A}'^{c}$. 

We define a distribution to be neutral if and only if the projections of $P$ on $\mathcal{A}'$ and $\mathcal{A}'^{c}$ are independent under the condition that any group of $P$ is either in $\mathcal{A}'$ or $\mathcal{A}'^{c}$. This condition is equivalent to having $P$ as the union of its two projections: $P=\pi(P) \cup \pi^{c}(P)$. A distribution is neutral if and only if:
\begin{gather}
    \textnormal{Pr}_{\alpha} \big( P = p \: | \: P = \pi(P) \cup \pi^{c}(P) \big) = \notag \\
    \textnormal{Pr}_{\alpha} \big( \pi(P) = \pi(p) \big) \times
    \textnormal{Pr}_{\alpha} \big( \pi^{c}(P) = \pi^{c}(p) \big).
\label{neutrality}
\end{gather}
We show in Appendix C that this property holds for any model specified for statistics $s_{k}$ defined as sums of real functions of the groups of $P$. Notably, all statistics proposed in the previous section and used in our analyses later are of this form, which allows us to interpret their associated parameters using simple log-odds ratios using the relations shown in \ref{subsec:Relations between partitions} (see Results section).

\section{Computation of the normalizing constant of the distribution}
\label{sec:Computation of the normalizing constant of the distribution}

\subsection{General case}
\label{subsec:General case}

As shown in the re-wiring of the Ewens formula in Equation \eqref{equ:Ewens}, some model specifications induce a simplification of Equation \eqref{equ:ERPM_jointform} into more tractable forms. This allows a direct evaluation of \eqref{equ:ERPM_jointform} for these cases, which can be leveraged to approximate the likelihood of more complex specifications, that we use in the empirical example of the paper.

For a model specified only with the statistic $s_{1}(P) = \#P$ for a set of $n$ nodes, we can make use of the Stirling numbers of the second kind \shortcite{riordan2012introduction} to derive a simple formulation of the normalizing constant $\kappa_{\mathcal{P}}$. The Stirling number ${n\brace m}$ is the number of partitions with $m$ groups, in other words, the number of partitions for which $s_{1}(P)=m$ \shortcite{pitman1997some}. We can therefore sum over all possible values $m$ and get the direct expression:
\begin{equation}
    \kappa_{\mathcal{P}}(\alpha_{1}) = \sum_{m = 1}^{n}{{n\brace m}\exp(\alpha_{1}m)}.
\label{equ:kappa_numgroups}
\end{equation}

More interestingly, one can calculate the normalizing constant of any model containing statistics of the form:
\begin{equation}
    s_{k}(P) = \sum_{G \in P}{f_{k}(\#G)}
\label{function_sizes}
\end{equation}
where $f_{k}$ are functions of the block sizes. For such models, the sufficient statistics define an exchangeable distribution \shortcite{mccullagh2011random} that does not depend on the labeling of the nodes. We define $\kappa_n$ as the normalizing constant of these models on any set of $n$ nodes. This constant can be constructed as a recursive sequence and computed with the following formulas:
\begin{gather}
    \kappa_{0} = 1, \; \kappa_{1} = \exp \Big( \sum_{k \in K}{\alpha_{k}f_{k}(1)} \Big), \notag \\
    \kappa_{n+1} 
    = \sum_{i=0}^{n}{ \binom{n}{i} \exp \bigg( \sum_{k \in K}{\alpha_{k} f_{k}(n+1-i)} \bigg)  \kappa_{i} }. 
\label{kappa_recurrence}
\end{gather}

The proof of the derivation of these relations can be found in Appendix D.

\subsection{Restriction to smaller supports}
\label{subsec:Restriction to smaller supports}

As mentioned in Section 2.3, some analyses might require restricting the sampled space of partitions. For the interest of this paper, we focus on the subset $\mathcal{P}'$ of partitions containing only groups of sizes between $\sigma_{min}$ and $\sigma_{max}$. 

The formula \eqref{equ:kappa_numgroups} previously established for the sole statistic $s_1(P) = \#P$ can also be used by replacing the Stirling numbers ${n \brace m}$ by an extension defined by the number of partitions in $m$ blocks with all block sizes belonging to $[\sigma_{min},\sigma_{max}]$. Appendix A provides details on how to recursively calculate these numbers.

More generally, the property given by formula \eqref{kappa_recurrence} can be extended for this case of size restrictions (see Appendix D). Again, models defined for statistics of the form \eqref{function_sizes} on the set $\mathcal{P}'$ are exchangeable and we can write the constant $\kappa_{\mathcal{P}'}$ as $\kappa_{n}'$ as it only depends on the number of nodes. By using again the values $i_{min} = \max(0,n+1-\sigma_{max})$ and $i_{max} = \min(n,n+1-\sigma_{min})$, we can construct the sequence $\kappa_{n}'$ with the following recursion:
\begin{gather}
    \kappa_{n}' = 0 \textnormal{ for } 0<n<\sigma_{min}, \notag \\
    \kappa_{0}' = 1, \; \kappa_{\sigma_{min}}' = \exp \Big(\sum_{k \in K}{\alpha_{k}f_{k}(\sigma_{min})} \Big), \notag \\
    \kappa_{n+1}' = \sum_{i=i_{min}}^{i_{max}}{ \binom{n}{i} \exp \bigg( \sum_{k \in K}{\alpha_{k} f_{k}(n+1-i)} \bigg)  \kappa'_{i} } \textnormal{ for } n \geqslant \sigma_{min}.
\label{equ:kappa_recurrence_restricted}
\end{gather}

\section{Estimation}
\label{sec:Estimation}

In the case of exponential families, the maximum-likelihood estimation method is equivalent to the method of moments that consists in finding the parameters under which the expected statistics of the modeled partition are equal to the observed statistics \shortcite{sundberg2019statistical}. Such estimations require, however, the calculation of either the likelihood function or the expected statistics under the model.
When the normalizing constant of a model, and therefore its likelihood, can be calculated as shown in Section \ref{sec:Computation of the normalizing constant of the distribution}, any optimisation method, such as a Newton-Raphson method \shortcite{deuflhard2011newton}, can be applied to approximate the parameter value for which this likelihood is maximum. This maximum either exists and is unique, or is infinite, by virtue of the properties of convexity in exponential families \shortcite{wedderburn1976existence}.

As soon as a model includes statistics related to actors' attributes, such simplifications of the normalizing constant $\kappa_{\mathcal{P}}$ are unlikely to be found. Since $\kappa_{\mathcal{P}}$ contains $B_{n}$ terms, following Equation \eqref{equ:bell_number}, the calculation of this likelihood is practically intractable for a large number of nodes. This problem can be circumvented with Monte-Carlo Markov Chain (MCMC) techniques, drawing inspiration from algorithms originally devised for ERGMs \shortcite{lusher2013exponential,snijders2002markov,hunter2006inference}.

\subsection{Sampling partitions}
\label{subsec:Sampling partitions}

As the space $\mathcal{P}$ becomes extremely large for high values of $n$, we can only sample a subset of random partitions to approximate the distribution of partitions under a given model. Monte-Carlo Markov Chain (MCMC) methods can assist in constructing such a subset by sampling partitions from a Markov chain whose stationary distribution is the model distribution given by Equation \eqref{equ:ERPM_jointform}. A suitable algorithm for this purpose is the Metropolis-Hastings algorithm \shortcite{metropolis1953equation, hastings1970monte, chib1995understanding}. 

The Metropolis-Hastings approach consists in defining a Markov chain with transition probabilities $q$ defined as the following product:
\begin{equation}
    q(p'|p)=\widetilde{q}(p'|p)A(p',p).
    \notag
\end{equation}
 At each step in the chain, a new partition $p'$ is proposed with probability $\widetilde{q}(p'|p)$, and it gets accepted according to the acceptance ratio $A(p',p)$. 

To define the proposal distribution $\widetilde{q}$, we use a symmetric relation $\mathcal{R}$ on the space $\mathcal{P}$. This relation can be one of the previously defined relations $\mathcal{R}^{\textit{merge}}$, $\mathcal{R}^{\textit{permute}}$, and $\mathcal{R}^{\textit{transfer}}$ or a combination of them. Importantly, this relation should connect the entire outcome space ; therefore, $\mathcal{R}^{\textit{permute}}$ should not be used without at least one of the other two, because it maintains the size structure. For our analyses, we use $\mathcal{R}^{\textit{merge}}$ for purely structural models, and a combination of the three relations when covariate effects are included. To know how often each relation should be used, one can try different combinations and pick the one that leads to the better mixed chain.

For a given partition $p$, we propose to only move to a partition $p'$ such that $p$ and $p'$ are linked by a given relation $\mathcal{R}$, with a uniform probability:
\begin{equation}
    \widetilde{q}(p'|p) =  
       \frac{1}{\#\{\widetilde{p} \: | \: (p,\widetilde{p}) \in \mathcal{R}\}}.
\notag
\end{equation}
By using the detailed balance equation $q(p'|p)\textnormal{Pr}(p) = q(p|p')\textnormal{Pr}(p')$ that ensures the convergence of the Markov chain to the desired distribution \shortcite{metropolis1953equation, hastings1970monte, chib1995understanding}, we get:
\begin{equation}
    A(p',p) = \textnormal{min} \Bigg( 1, \frac{\textnormal{Pr}(P') \; \#\{\widetilde{p} \: | \: (p,\widetilde{p}) \in \mathcal{R}\} }
    {\textnormal{Pr}(P)\; \#\{\widetilde{p} \: | \: (p',\widetilde{p}) \in \mathcal{R}\} } \Bigg).
\notag
\end{equation}

As evident from Figure \ref{fig:relation_illustration}, the proposal distribution defined for relations such as $\mathcal{R}^{\textit{merge}}$ is not symmetric. In other words, for some pairs $(p,p') \in \mathcal{R}$ it is the case that $\widetilde{q}(p'|p) \neq \widetilde{q}(p|p')$. Therefore, it is necessary to calculate the proposal probabilities at each step of the chain to find the acceptance ratio. Deciding on which relation to use depends on how fast these calculations can be made and on how efficiently the algorithm covers the sampled space. Moreover, the proposal distribution has to be adapted when the set of allowed partitions is restricted, to make sure only and every correct partition is reached. In certain cases, it might be well advised to design a chain that covers a larger space and only retain correct partitions.

\subsection{Estimation procedure}
\label{subsec:Estimation procedure}

The estimation procedure used in this study implements the Robbins-Monro algorithm \shortcite{robbins1951stochastic} in a similar way as used in \citeauthor{snijders2002markov} \citeyear{snijders2001statistical,snijders2002markov} for the estimation of ERGMs. The Robbins-Monro algorithm is a variant of the Newton-Raphson optimisation algorithm for objective functions obtained via Monte Carlo methods. It was shown to be a useful tool for a large range of stochastic approximation problems \shortcite{lai2003stochastic}, in particular for the maximum-likelihood estimation of models that can only be analyzed by simulations \shortcite{cappe2005inference, gu1998stochastic, gu2001maximum}. Although this algorithm was chosen for our study, we note that various other algorithms were designed for similar problems, among which notably the Geyer-Thompson algorithm \shortcite{geyer1992constrained} and the stepping algorithm by Hummel and colleagues \citeyear{hummel2012improving}.


The aim of the Robbins-Monro algorithm is to solve the moment equation:
\begin{equation}
    E_{\alpha}[s] = s_{obs},
    \label{eq:optim_problem_RM}
\end{equation}
where $E_{\alpha}[s]$ is the expected vector of sufficient statistics for the model with parameter $\alpha$ and $s_{obs} = s(p_{obs})$ is the vector of statistics in the observed partition $p_{obs}$. The original $N^{th}$ iteration step of the algorithm consists in drawing a variable $s_{N}$ from the distribution of the statistics for the model with parameter $\alpha_{N}$ and updating the model parameter to:
\begin{equation}
    \alpha_{N+1} = \alpha_{N} - a_{N} D_{N}^{-1} (s_{N} - s_{obs}).
    \label{eq:update_RM}
\end{equation}
In this equation, $(a_{N})$ is called the \textit{gain sequence} and controls the magnitude of the optimisation steps and $D_{N}$ is the \textit{scaling matrix}. A classic choice for the gain is $a_{N} = 1/N$ and for $D_{N}$ the derivative matrix ${\partial E_{\alpha_{N}}[s]}/{\partial \alpha_{N}}$. 

Using the arguments developed by \citeauthor{snijders2001statistical} \citeyear{snijders2001statistical}, our algorithm uses in place of the matrices $D_{N}$ only one scaling matrix $D_{0}$ calculated once for all. This scaling matrix is the covariance matrix of a sample of the model parametrized by some starting parameters, and represents an estimation of the sensitivity of the sufficient statistics to the parameters' variations. 
This is based on a result from \citeauthor{polyak1990new} \citeyear{polyak1990new} implying that the use of this matrix, or also its diagonal matrix, will lead to an optimal rate of convergence, as long as the sequence $(a_{N})$ converges at the rate $N^{-c}$, with $0.5 < c < 1$. This procedure also requires to use the tail average of the sequence $(\alpha_{N})$ as the solution to the optimisation problem \eqref{eq:optim_problem_RM}.

Regarding the gain sequence, we use the idea from \citeauthor{pflug1990non} \citeyear{pflug1990non} that it is better to keep a constant value $a_{N}$ as long as the sequence $s_{N}$ has not crossed the observed values $s_{obs}$ yet. The algorithm is therefore divided in $R$ subphases within which the value $a_{r}$ is kept constant while the sequence $(\alpha_{r,N})$ is updated with the adapted steps \eqref{eq:update_RM}:
\begin{equation}
    \alpha_{r,N+1} = \alpha_{r,N} - a_{r} D_{0}^{-1} (s(p_{N}) - s_{obs}),
    \label{eq:update_adapted_RM}
\end{equation}
with $p_{N}$ drawn from the model parametrized by $\alpha_{r,N}$.
Importantly, the lengths of the subphases must ensure the convergence of $(a_{N})$ at the rate $N^{-c}$, and the starting parameter value for the subphase $r$ should be the average of the previous sequence $(\alpha_{r-1,N})$, in order to satisfy the convergence conditions mentioned earlier.

In practice, the algorithm is implemented in three phases. The first phase is used to estimate the matrix $D_{0}$ by sampling $M_{1}$ partitions $p_{1}$, $p_{2}$, ..., $p_{M_{1}}$ from the model defined for the starting parameters $\alpha_{0}$, with the Metropolis-Hastings algorithm presented in Section \ref{subsec:Sampling partitions}. A good choice for $\alpha_{0}$ usually is a vector containing zeros except for parameters that can be calculated with the equations shown in Equation \ref{sec:Computation of the normalizing constant of the distribution}. We only retain partitions after a \textit{burn-in} period and with a certain \textit{thinning} interval as to ensure a low auto-correlation between the sampled statistics (below 0.4 is an efficient rule of thumb). A value of a few hundreds for $M_1$ usually sufficed. We obtain an estimation of the expected statistics and of the covariance matrix:
\begin{equation}
    \overline{s}_{\alpha_{0}} = \frac{1}{M_{1}}(s(p_{1}) + s(p_{2}) + ... + s(p_{M_{1}}))
    \notag
\end{equation}
\begin{equation}
    \hat{\textnormal{cov}}(\alpha_{0}) = \frac{1}{M_{1}}\sum_{m=1}^{M_{1}}{(s(p_{m}){s(p_{m})}^{T})} - \overline{s}_{\alpha_{0}}{\overline{s}_{\alpha_{0}}}^{T}
    \notag
\end{equation}
The scaling matrix $D_{0}$ is defined\footnote{To achieve a more stable algorithm, the non-diagonal elements of $D_0$ can be multiplied by a constant between 0 and 1 (in our examples, $0.2$ was used).} defined as $D_{0} = \textnormal{diag} \big( \hat{\textnormal{cov}}(\alpha_{0}) \big)$. Its inverse $D_{0}^{-1}$ provides the new starting estimates:
\begin{equation}
    \alpha_{0} - a D_{0}^{-1} (\overline{s}_{\alpha_{0}} - s_{obs}).
    \notag
\end{equation}

In the second phase, we implement the iterative steps of \eqref{eq:update_adapted_RM} within $R$ subphases. At each $N^{th}$ iteration, only one partition $p_{N}$ is drawn from the distribution with parameter $\alpha_{r,N}$, with the Metropolis-Hastings algorithm starting at the previously drawn partition $p_{N-1}$. Each $r$ subphase lasts until its length is above the minimum length of the subphase and all sampled statistics have crossed the observed values. Alternatively it stops when $N$ is above the maximal length of the subphase. In this study, we used the values $R=4$, $a=0.1$, $a_{r} = a / (2^{r-1})$, and kept lengths of subphases of the order $2^{4r/3}$ that ensured the crossing of statistics.

Finally, phase 3 is used to sample $M_3$ partitions from the final distribution in order to approximate the expected sufficient statistics with the sample mean $\overline{s}_{\alpha_{f}}$ and the covariance matrix of these statistics with the sample covariance matrix. We used large values of $M_3$, typically between $1000$ and $2000$. Model convergence is assessed by calculating the sample standard deviation for each statistic separately. It is considered excellent for the $k$th statistic when the convergence ratio $c_{k}$:
\begin{equation}
 c_{k}= \frac{\overline{s}_{\alpha_{f},k} - s_{obs,k}}{\textnormal{SD}_{\alpha_{f}}(s_{k}(p_{1}),...,s_{k}(p_{M_3}))}
 \notag
\end{equation}
remain between $-0.1$ and $0.1$, with $\textnormal{SD}_{\alpha_{f}}(s(p_{1}),...,s(p_{M_3}))$ being the sample standard deviations. This value is aligned to the one chosen for ERGM estimation (see \shortcite{snijders2002markov}). Furthermore, we assume that parameter estimates have an approximate multivariate normal distribution, similarly to ERGMs \shortcite{lusher2013exponential}. We can therefore test significance of the model parameters from a simple Wald test considering whether the ratio between the elements of $\alpha_{f}$ and their standard errors (calculated from the inverse of the sample covariance matrix) are smaller than $-2$ or larger than $2$. 

\subsection{Model diagnostics}
\label{subsec:Model diagnostics}

The goodness of fit of a model can be assessed by the calculation of auxiliary statistics (i.e., not included in the sufficient statistics of the model), similarly to ERGMs \shortcite{hunter2008goodness}. By sampling from the estimated model, we can test whether the obtained distribution of such auxiliary statistics correspond to those in the observed data. 

In order to compare different model specifications, we further calculate log-likelihoods, in a similar way to the one proposed by \citeauthor{hunter2006inference} \citeyear{hunter2006inference} for ERGMs. This calculation is done through path-sampling as presented by \citeauthor{gelman1998simulating} \citeyear{gelman1998simulating} to estimate the log-likelihood of a model for an estimated parameter $\alpha$ when its normalizing constant $\kappa(\alpha)$ is intractable. First, we calculate the log-likelihood $\ell(\alpha_{0},p_{obs})$ of the model parametrized by $\alpha_{0}$ in which statistics are identical to the ones in the model of interest but all parameters except the one for the statistic $s_{1}(P) = \# P$ are set to zero. The value of $\alpha_{0}$ is calculated using the equations of Section \ref{sec:Computation of the normalizing constant of the distribution}.
We can then estimate the difference between the normalizing constants $\lambda(\alpha_{0},\alpha) = \kappa(\alpha) - \kappa(\alpha_{0})$ by sampling $M$ models with parameters $\alpha_{m} = \frac{m}{M}\alpha + \frac{1-m}{M}\alpha_{0})$ that produce large overlaps between the sampled distributions:
\begin{equation}
    \hat{\lambda}(\alpha,\alpha_{0}) = \frac{1}{M}\sum_{m=1}^{M}{(\alpha - \alpha_{0})^{T}\overline{s}_{\alpha_{m}}}.
    \notag
\end{equation}
We finally estimate the log-likelihood of our model of parameter $\alpha$ with:
\begin{equation}
    \hat{\ell}(\alpha,p_{obs}) = \ell + (\alpha - \alpha_{0})^{T} s_{obs} - \hat{\lambda}(\alpha,\alpha_{0}).
    \notag
\end{equation}
Implemented code, documentation (including UML charts), an example script, and the script used for all the presented analyses can be found in the supplementary materials and the repository \textit{github.com/marion-hoffman/ERPM}.
The results of the Robbins-Monro algorithm were compared to a simple Newton-Raphson estimation in the case of a simple model for which the likelihood can directly be calculated.

\section{Case study: the composition self-formed teams during hackathons}
\label{sec:Case study}


\subsection{Data}
\label{subsec:Data}

Hackathons were defined as "problem-focused computer programming events" by \citeauthor{topi2014computing} \citeyear{topi2014computing}. They are often designed for participating teams to solve a digital problem in a short period of time. Such events provide companies, universities, or non-profit organizations the opportunity of harnessing the ideas of volunteers in exchange for rewards and funding for the winning teams \shortcite{lara2016hackathons,briscoe2014digital}. Hackathons have recently developed to tackle an increasingly broad range of topics, including education, marketing, and arts \shortcite{lara2016hackathons}.

We collected data during two editions of a hackathon at a technical university. The events welcomed $60$ and $58$ participants respectively, who divided themselves into $14$ teams in both cases. Individual attributes of participants as well as their prior acquaintances were gathered during the registration process via online questionnaires. The events were scheduled as follows. The registered participants were invited to the venue on a Saturday at 9:00 and were introduced to the tasks proposed to them. They were later asked to mingle and define teams until 13:00. Organizers only allowed teams including $2$ to $5$ individuals in the first edition and $3$ to $5$ members in the second. 
These teams collaborated until Sunday afternoon on designing and implementing their solution to the hackathon challenge. 
The teams' compositions and their performances as assessed by a jury of experts were collected at the end.

In the first edition, $1$ team of $2$, $1$ team of $3$, $5$ teams of $4$, and $7$ teams of $5$ were formed. The $14$ teams in the second edition were divided into $1$ team of $3$, $9$ teams of $4$, and $3$ teams of $5$.
Descriptives for the participants' attributes used in our analyses are presented in Table \ref{tab:descriptives}. Additionally, $22$ pairs of participants reported already knowing each other in the first edition, and $23$ such pairs were reported in the second edition.

\begin{table}[h]
   \centering
   \small
\renewcommand{\arraystretch}{1.2}
\begin{tabular}{l@{\extracolsep{0.5cm}}l@{\extracolsep{0.5cm}}  
c@{\extracolsep{0.5cm}}c@{\extracolsep{0.5cm}}
c@{\extracolsep{0.5cm}}c@{\extracolsep{0.5cm}} l}

&& \multicolumn{2}{c}{\textbf{First edition}}  & \multicolumn{2}{c}{\textbf{Second edition}}  \\  \hline
    
    \textbf{Gender} & Male & ($N=60$) & 49 & ($N=58$) & 55 \\
    & Female &  & 11 & & 3  \\ \hline
	
    \textbf{Age} & $<$ 20 & ($N=43$) & 11 & ($N=54$) & 12\\
    & 20-25 &  & 13 & & 25 \\ 
    & 25-30 &  & 10 & & 13 \\ 
	& $>$ 30 &  & 9 & & 4 \\ \hline
	
	\textbf{First language} & Swiss German & ($N=49$) & 16 & ($N=56$) & 16  \\
    & German & & 10 & & 10 \\
    & Others & & 23 & & 30 \\ \hline
	
  
    \textbf{Major} & Engineering & ($N=60$) & 14 & ($N=58$) & 34 \\
    & Computer Science, IT & & 23 & & 10 \\ 
    & Physics & & 6 & & 3 \\  
    & Mathematics & & 2 & & 2 \\  
    & Chemistry & & 3 & & 5 \\  
    & Environmental sciences & & 4 & & 3 \\  
    & Other & & 8 & & 1 \\  
    
    \end{tabular}
    \caption{Counts of gender, age, language, degree, and major attributes among participants of the first and second hackathon editions.}
    \label{tab:descriptives}
\end{table}

\subsection{Theoretical mechanisms of team formation}
\label{subsec:Theoretical mechanisms of team formation}

Self-assembled teams for short projects are ubiquitous to organizational, educational, or recreational contexts \shortcite{falk2010advancing, guimera2005team, contractor2013some, zhu2013motivations}. 
Scholars investigating the motivations for individuals to form a team in various settings generally identify four types of mechanisms as classified by \shortcite{bailey2017social}, namely, homophily, competence, familiarity, and affect.

First, homophily, as reviewed by \shortcite{mcpherson2001birds}, is commonly observed in dyadic collaboration and teams \shortcite{kalleberg1996organizations, ruef2003structure,mcpherson1987homophily, gompers2017homophily}, and denotes that similar individuals tend to collaborate. 
In our context, being of a similar age or sharing the same language could enable communication within the teams and could have therefore contributed to the choice of teammates. Other attributes, such as gender or personality traits could have also been relevant for this context but were discarded because of a lack of variation or simply because of a lack of explanatory power of the models.

Second, the competence of team members is central in teams whose aim is to achieve a given task. 
The difficult endeavor of forming performing teams is to find the right balance between optimizing the number of skills within team members and reducing overhead costs of combining different ways of thinking or working. 
Previous research on self-assembled teams found evidence for complementarity of skills \shortcite{zhu2013motivations}, as well as skill homophily \shortcite{gomez2019would}. 
During the hackathons, organizers strongly recommended participants to form teams with as diverse skills and knowledge as possible. Consequently, we tested whether participants were more likely to form teams with individuals coming from different majors. 

Third, familiarity describes that individuals are more comfortable teaming up with others with whom they have collaborated in the past, because of shared practices or values  \shortcite{bailey2017social,lungeanu2018team,gomez2019would}. 
Since some participants knew each other prior to the event and reported participating together, we expected to find a high number of prior acquaintance ties within the teams.

Finally, interpersonal affect, or on the opposite dislike, can be a strong predictor in the choice of team partners, arguably even more important than competence \shortcite{casciaro2008competence}. However, since our data did not contain such information, we did not test any related mechanism.

\subsection{Results}
\label{subsec:Results}

Three models are presented for each dataset. The first two models include effects related to group sizes and individual attributes, and are used to explore the specification of statistics related to age and language. The third models are shown as final models and include the additional effect of previous acquaintances. All models are reported in Table \ref{tab:results201718}.

The group size distribution is modeled by the number of groups and the sum of squared sizes in the first dataset. In the second dataset, the second statistic is excluded because of degeneracy issues (see Section \ref{subsubsec:Structural statistics}) and replaced by the number of groups of size $4$ (since participants were advised, although not obligated, to form specifically groups of this size). In both cases, we limit the allowed group sizes to a minimum of $2$ and a maximum of $5$ in the estimation procedure. 
Regarding attribute effects, three homophily effects are included for age, language, and major. Age homophily is modeled using the sum of age ranges in the groups (except in Model 1 of the first dataset). Homophily for language and major uses the sum of different attributes present in each group (except in Model 1 of the second dataset in the case of language). Finally, the influence of previous acquaintances is captured by the count of such ties within all teams.

\begin{table}[b!]
   \centering
   \small
\renewcommand{\arraystretch}{1.2}
\begin{tabular}{ l@{\extracolsep{0.2cm}} l@{\extracolsep{0.2cm}}  
r@{\extracolsep{0.1cm}}c@{\extracolsep{0.1cm}}l@{\extracolsep{0.2cm}}
r@{\extracolsep{0.1cm}}c@{\extracolsep{0.1cm}}l@{\extracolsep{0.2cm}}
r@{\extracolsep{0.1cm}}c@{\extracolsep{0.1cm}}l@{\extracolsep{0.2cm}}}

&& \multicolumn{3}{c}{\textbf{Model 1}}  & \multicolumn{3}{c}{\textbf{Model 2}} & \multicolumn{3}{c}{\textbf{Model 3}}  \\  
& Sufficient statistic & Est. & Sig. & S.e.  & Est. & Sig. & S.e. & Est. & Sig. & S.e.  \\ \hline \hline
    
\textbf{First edition} & Number of groups & --4.67 & & (4.98) & --4.73 & & (4.85) & --4.62 & & (4.73) \\
    & Sum of squared sizes & 0.05 & & (0.35) & --0.06 & & (0.34) &  --0.07 &  & (0.34) \\ \cline{2-11}
    
    & Age differences &  --0.027 &  & (0.021) &  &  &  &  &  &  \\ 
    & Age ranges &  &  &  & --0.16 & * & (0.08) & --0.10 &  & (0.08) \\ 
    & Number of languages & --0.10 & & (0.51) & --0.09 &  & (0.51) & 0.29 & & (0.50) \\ 
    & Number of majors & 0.05 & & (0.47) & 0.02 &  & (0.48) & 0.29 & & (0.50) \\ \cline{2-11}
    
    & Number of acquaintances &  &  &  &  &  &  & 2.90 & *** & (0.46) \\ \hline
    
    Log-likelihood && \multicolumn{3}{c}{--124.2} & \multicolumn{3}{c}{--123.7} & \multicolumn{3}{c}{--102.1} \\ \hline \hline

\textbf{Second edition} & Number of groups & --4.00 & & (2.44) & --3.85 &  & (2.65) & --3.74& & (2.44) \\
& Number of groups size 4 & 2.02 & *** & (0.59) & 2.03 & *** & (0.61) & 2.02 & ***& (0.62) \\\cline{2-11}
    
    & Age ranges & --0.29 & * & (0.12) & --0.27 & * & (0.12) & --0.18 & & (0.12) \\ 
    & Same language pairs & 0.72 & *** & (0.21) & &  &  &  &  &  \\ 
    & Number of languages &  &  &  & --1.54 & *** & (0.43) & --1.23 & ** & (0.46) \\ 
    & Number of majors & --0.33 & & (0.58) & --0.49 & & (0.61) & --0.07 & & (0.60) \\ \cline{2-11}
    
    & Number of acquaintances &  &  &  &  &  &  & 2.41 & *** & (0.39) \\ \hline
    
    Log-likelihood && \multicolumn{3}{c}{--106.8} & \multicolumn{3}{c}{--105.9} & \multicolumn{3}{c}{--89.8} \\
    
    \end{tabular}
    \caption{Estimated parameters for the models of the two hackathon editions. Convergence ratios for each parameter are below $0.12$.}
    \label{tab:results201718}
\end{table}

In Models 1 and 2 for the first dataset, we compare two specifications for age homophily, using respectively the sum of absolute differences in age among teammates (i.e, dyadic homophily effect) and the sum of age ranges in each team (i.e., group homophily effect). In both models, the age-related parameter is negative, indicating a tendency to form groups with low age differences. However, only the parameter of Model 2 is significant and the log-likelihood of $-124.2$ in Model 1 is also higher than the one of $-123.7$ Model 1 (these log-likelihoods can directly be compared because the number of parameters is constant across models). These points altogether suggest that the age range specification might better explain the data at hand. 

A similar exploration is carried out for the language specification in Models 1 and 2 of the second dataset, where language homophily is either specified by the number of same language ties within teams or the number of different languages in each team. This time, both parameters are strongly significant, and respectively indicate a tendency to form groups with a high number of same language ties or a small number of different languages. However, the slightly better log-likelihhood of Model 2 ($-105.9$ instead of $-106.8$) indicates that the second specification explains better these particular data.

We finally turn to the interpretation of the final models (Models 3). We first discuss one-by-one the direction and significance of the parameters. Interpreting the size of parameter values beyond its sign follows the same principles as interpretations for other exponential family models. We can use the binary relations \textit{merge/split}, \textit{permute}, and \textit{transfer} introduced in Section \ref{subsec:Relations between partitions} to define pairs of partitions that exhibit a unit change for a given statistic, \textit{ceteris paribus}. Using these operations we can formulate log probability ratios between partitions that are related through one of those relations and attach a quantitative interpretation to exact parameter values.

In the first dataset, the negative parameter for the number of groups of $-4.62$ indicates a tendency to form fewer and, therefore, larger groups. 
Here, we can interpret this parameter with the log probability ratio between two partitions linked by the \textit{merge/split} relation. Specifically, Model 3 predicts a partition to be around $\textnormal{exp}(4.62) \approx 101$ times more likely compared to the same partition with one group split into two, given that all other statistics remain constant and group sizes stay in the allowed range. This applies, for example, to the comparison of having one group of four participants compared to two groups of two participants, \textit{ceteris paribus}.
The negative parameter for squared sizes of $-0.07$ shows a concentration of sizes around large sizes (i.e., $4$ and $5$). However, these two effects are not significant (potentially because of the small size of the dataset). 
All effects related to individual attributes are also insignificant (including the effect of age that becomes explained away by the effect of previous acquaintances from Model 2 to 3). Their directions however suggest a tendency to form groups with low age differences, diverse languages, and diverse majors. 
Finally, we find a positive and strongly significant effect of previous acquaintances.
For this statistic, it is more useful to invoke the \textit{permute} relation to calculate log probability ratios.
Its parameter indicates that a partition obtained from a permutation of two actors that would add one acquaintance tie in a group, leaving other statistics equal, is around $\textnormal{exp}(2.90) \approx 18$ times more likely than before permutation.

Regarding the second dataset, the negative parameter for the number of groups of $-2.04$ in Model 3 shows a tendency to form fewer groups, but this effect is not significant either. The significant parameter of $2.02$ for groups of size $4$ indicates however a tendency of individuals to form more groups of this size than others.
The directions of the parameters for attribute-related effects suggest a tendency to form groups with low age differences and diverse majors, as in the other dataset, but with a low diversity in terms of languages. This significant effect for languages can be interpreted again with a log-odds ratio: A partition reached through the \textit{transfer} of an actor that would add a new language to a group is $\textnormal{exp}(1.23c) \approx 3.4$ times less likely than the partition before this transfer. 
Once again, the effect of previous acquaintances is strongly significant, with a similar magnitude as in the first dataset.

All in all, individual attributes seemed to have little influence on the composition of teams during the first edition. In particular, we do not find evidence for any homophily effect and the parameter related to majors even suggests that participants tried to diversify their teams as the organizers recommended (although this effect is not significant). On the other hand, we see that forming teams with similar languages was probably important during the second edition. For both editions, previous acquaintances seem to have been the strongest driver of team formation.

It is important to note that the probabilities in the log probability ratios mentioned above factorize for the models presented here because theses models respect the neutrality property defined by Section \ref{subsec:Independence properties of the distribution}. This means that the impact of the change of a statistic between two partitions can be interpreted net of the groups that are exactly equal between the two partitions. 
However, the \textit{ceteris paribus} condition is still not trivial to invoke, since it is not always possible to find partition changes that only affect one statistic at a time. Such log probability ratios should therefore be interpreted with caution.

\subsection{Model fit}


In this section, we further investigate the distribution of auxiliary statistics in the estimated models to assess goodness of fit, following similar procedures as the ones recommended for network models \shortcite{hunter2008goodness}. These distributions are represented by the violin plots proposed by \citeauthor{hintze1998violin} \citeyear{hintze1998violin}.

\begin{figure}[t]
    \centering
    \vspace{10pt}
    \includegraphics[scale=0.65]{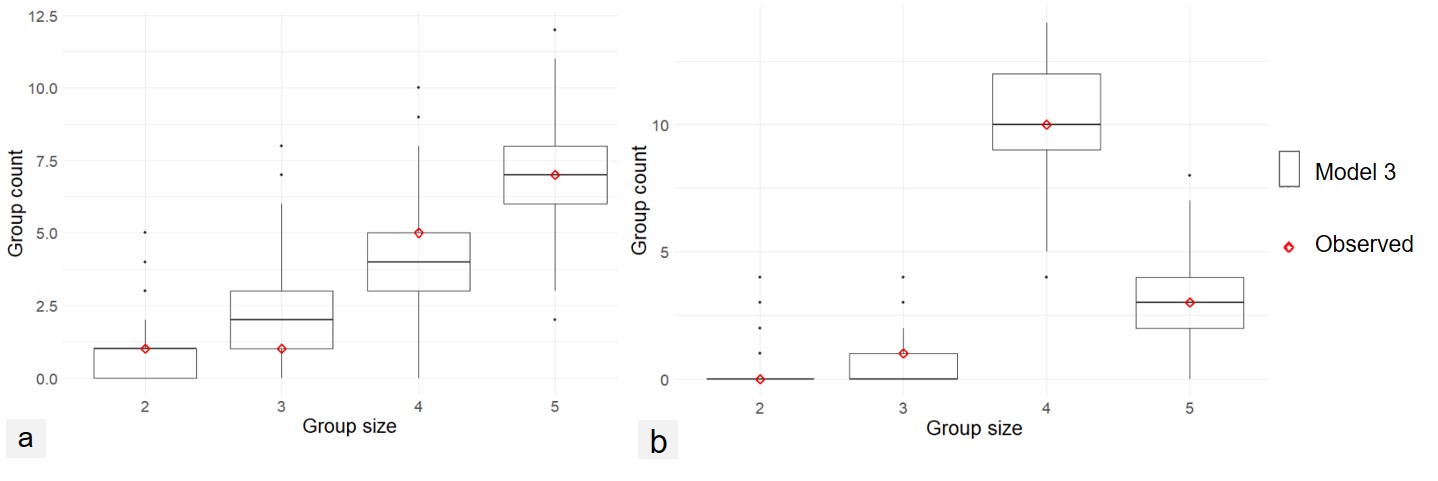}
    \caption{Group size distributions from Models 3 of the first (a) and second (b) datasets.}
    \label{fig:GOFs-sizes}
    \vspace{5pt}
\end{figure}

We first examine in Figure \ref{fig:GOFs-sizes} the distribution of group sizes in the final models (Models 3). We see in Figure \ref{fig:GOFs-sizes}a that the distribution of group sizes is well recovered for the first dataset, with the observed counts falling within the confidence intervals of the simulated models with a slight overestimation of groups of $3$ and a slight underestimation of groups of $4$. Sizes are also well recovered in the second dataset (\ref{fig:GOFs-sizes}b), for which the number of groups of size $4$ is perfectly estimated as it was a sufficient statistic of the model.

\begin{figure}[t!]
    \centering
    \vspace{10pt}
    \includegraphics[scale=0.75]{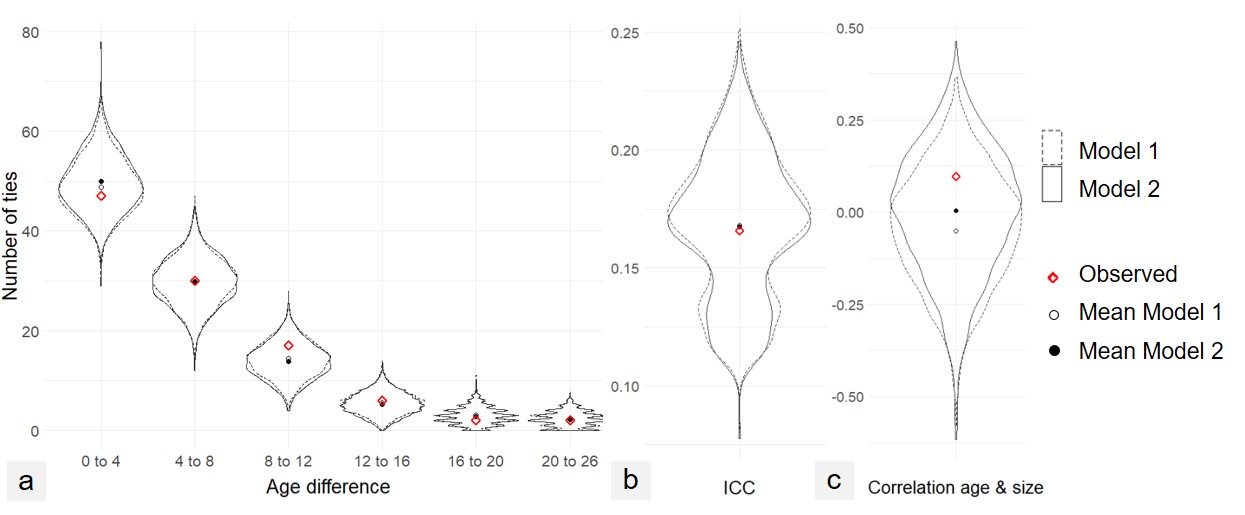}
    \caption{Distribution of auxiliary statistics related to age in simulated partitions from Models 1 and 2 for the first dataset: (a) number of ties with given age differences; (b) intraclass correlation coefficient for age; (c) Correlation between actors' age and their group size.}
    \label{fig:GOFs-age}
    \vspace{5pt}
\end{figure}

To understand better the specification of age homophily, we examine in Figure \ref{fig:GOFs-age} statistics related to the distribution of ages within the teams and compare their observed values to the values simulated by Models 1 and 2 for the first dataset. First, Figure \ref{fig:GOFs-age}a shows that both proposed specifications recover equally well the distribution of age difference within ties. To assess the homogeneity of groups in terms of age, we further calculate the intraclass correlation coefficient of ages within groups (Figure \ref{fig:GOFs-age}b). Again, both models provide a very satisfactory fit. Finally, examination of the correlation between individuals' ages and the size of their teams (Figure \ref{fig:GOFs-age}c) helps assessing how well the models reproduce the tendency of certain ages to be present in larger groups, which is not an effect included in the model. This shows that Model 2 provides a slightly better fit than Model 1, which might explain the better log-likelihood previously mentioned. This observation could be interpreted as the result of the first specification (sum of age differences) being dependent on group sizes while the second (sum of age ranges) is not, but this interpretation might only be valid for this particular data.

\begin{figure}[b!]
    \centering
    \vspace{10pt}
    \includegraphics[scale=0.73]{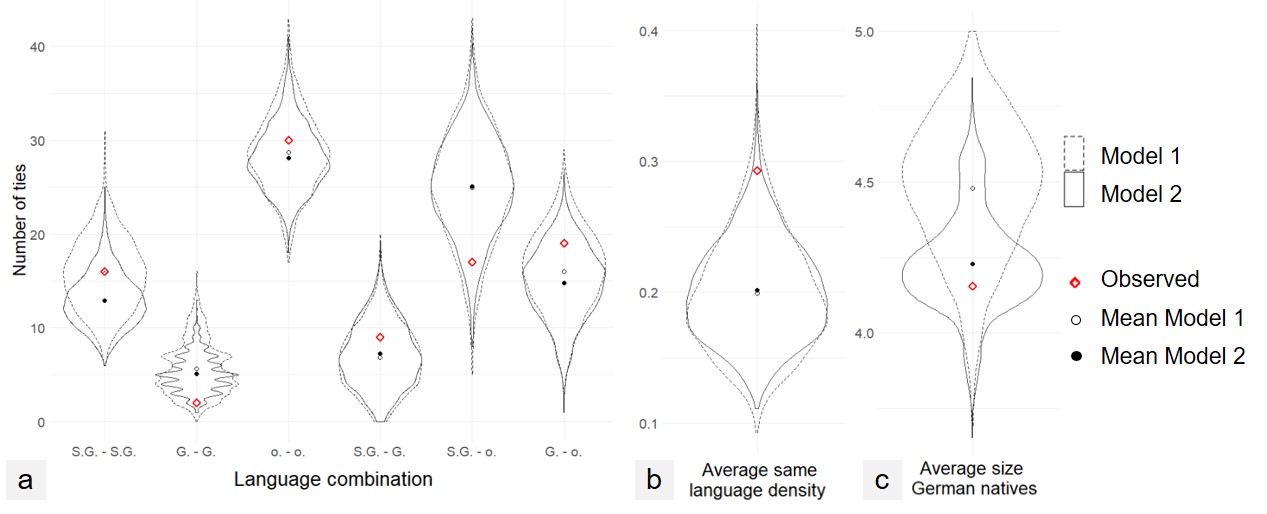}
    \caption{Distribution of auxiliary statistics related to language in simulated partitions from Models 1 and 2 for the second dataset: (a) number of ties with a given lamguage combination; (b) proportion of same-language ties; (c) Average group size of a native German speaker.}
    \label{fig:GOFs-language}
    \vspace{5pt}
\end{figure}

Moving on the specification of language homophily, Figure \ref{fig:GOFs-language} presents a similar type of auxiliary statistics, this time adapted to the categorical nature of the language attribute, for Models 1 and 2 of the second dataset. We first see in Figure \ref{fig:GOFs-language}a that the number of ties with certain language combinations are rather well reproduced by the models, with the exception of ties between Swiss German speakers and other language speakers. The homogeneity of groups is assessed by the proportion of same-language ties within groups (Figure \ref{fig:GOFs-age}b). The fit of this statistic is less satisfying, but equally so for both models. Finally, the average group size of a native German speaker is shown in Figure \ref{fig:GOFs-language}c to understand the link between language and group sizes. Again, Model 2 provides a better fit than Model 1, which could explain the better log-likelihood of the model.

\section{Conclusion}
\label{sec:Conclusion}

The present paper introduces the statistical framework of exponential partition models and presents its main mathematical properties. Building upon the rich literature on exponential families of distributions, stochastic networks, and stochastic partitions, we show that this model can uncover regularities in observations of self-assembled exclusive groups while taking into account structural dependencies between these observations. 
Exponential partition models can be applied to various contexts in which individuals sort themselves into groups based on social preferences, opportunities, and exogenous constraints. 
Specifications are proposed to investigate a variety of mechanisms that can be situated at an individual, relational, and group level. An example study case illustrating some of the capabilities of the model using the self-formation of hackathon teams is provided. 
All code and documentation for further use of this framework can be found at \textit{github.com/marion-hoffman/ERPM}. 
Data for replication are available on request.

This work bridges two branches of the statistics literature, one representing systems as networks and the other as partitions. 
On the one hand, we augment network methods by introducing the possibility of modeling social mechanisms at the level of groups rather than dyads. 
By re-thinking the mathematical representation of groups, the proposed framework allows researchers to investigate group formation as coordination processes between individuals rather than an aggregation of dyadic ties to group entities. 
On the other hand, we contribute to the stochastic partition modeling literature by extending the use of such models to studying complex structural properties of social communities. 
In particular, the model allows to study the influence of mechanisms related to individual and relational covariates on group formation processes.  

The presented methodological developments aim to further our understanding of mechanisms driving the formation of social groups. First, they allow social scientists to model and explain observations of self-assembled groups and potentially expand the range of social contexts that could be investigated. 
Moreover, some social processes widely studied at the dyadic level, such as homophily, can now be investigated at the group level. 
This modeling framework offers the possibility to explore different operationalizations of such mechanisms and assess which ones give a better representation of real-life processes, through the use of model diagnosis techniques described in this paper. 
Finally, by moving from the dyad to the group perspective, mechanisms that have been suggested for group processes can be statistically tested. 

Much remains to be discovered about the formation of self-assembled groups. 
A current limitation of the presented framework is its inability to model observations of overlapping groups. 
Since such groups are encountered in many social contexts, future research should extend the modelling framework to more general data representations, such as hypergraphs. 
Modeling group overlaps opens up the possibility of representing new dependencies between group memberships and to analyse, for example, what leads individuals to belong to multiple groups at the same time. 
A second limitation lies in its cross-sectional nature. 
Dynamic or longitudinal data offering rich insights on the processes driving social systems, an extension of this framework to a dynamic group representation would greatly further our understanding of social groups dynamics.


\section*{Acknowledgement(s)}
The authors thank the members of the Chair of Social Networks at ETH Zürich, and the members of the Duisterbelt, for useful comments and feedbacks. P.B. was supported by the Leverhulme Centre for Demographic Science.

\section*{Disclosure statement}
The authors report no conflict of interest.

\bibliographystyle{apacite}
\bibliography{bibliography}

\newpage

\noindent\textbf{Appendix A. Partition sets with restricted group sizes: extension of the Bell numbers and Stirling numbers of the second kind}

\bigskip

\noindent \textbf{Extended Bell numbers.}
Bell numbers can famously be derived through the recursive relation \eqref{equ:bell_number} \shortcite{bell1934exponential}, and a similar formula can express the size of the space $\mathcal{P}'$ containing all partitions whose group sizes belong to the interval $[\![\sigma_{min},\sigma_{max}]\!]$. 

To initialize the recurrence, we first know that the sets $\mathcal{P}'([\![ 1,n ]\!])$ are empty when $n$ is smaller then $\sigma_{min}$. The minimal size $n$ required for one correct partition to exist, therefore we have:
\begin{equation}
    B'_{n} = 0 \textnormal{ for } 0< n < \sigma_{min} \textnormal{ and }
    B'_{\sigma_{min}} = 1 \notag 
\end{equation}

The Bell recursion at $(n+1)$ formula enumerates, for $i$ varying from $0$ to $n$, the partitions with node $(n+1)$ in a group of size $(n+1-i)$ and the $i$ remaining nodes covering all possible partitions given by $B_{i}$. Here, we can enumerate the same partitions but the size $(n+1-i)$ can only take values between $\sigma_{min}$ and $\sigma_{max}$, therefore $i$ can only vary from $i_{min}$ and $i_{max}$ defined as:
\begin{equation}
    i_{min} = max(0, n+1-\sigma_{max}) \textnormal{ and } i_{max} = min(n, n+1-\sigma_{min}).
\label{iminmax}
\end{equation}
If we note $\mathcal{P}_{i}([\![ 1,n+1 ]\!])$ the sets containing all partitions $P \in \mathcal{P}'([\![ 1,n+1 ]\!])$ such that node $(n+1)$ belongs to a group of size $(n+1-i)$:
\begin{equation}
    \mathcal{P}_{i}([\![ 1,n+1 ]\!]) = \{ P \in \ \mathcal{P}'([\![ 1,n+1 ]\!]) \; | \; \#g_P(n+1) = n+1-i \},
\label{subsetsi}
\end{equation} 
we can write:
\begin{equation}
    \mathcal{P}'([\![ 1,n+1 ]\!]) = \bigcup_{i=i_{min}}^{i_{max}}{\mathcal{P}_{i}([\![ 1,n+1 ]\!])}
\notag
\end{equation}

For partitions in $\mathcal{P}_{i}([\![ 1,n+1 ]\!])$, we first know that there are $\binom{n}{i}$ ways to choose the group of $(n+1)$ and $B'_{i}$ ways to choose how to arrange the remaining $i$ nodes. This is true for any $i$ except when $(n+1-i)$ corresponds to the whole set size $(n+1)$ (i.e., $i=0$). In that case, we use for convenience:
\begin{equation}
    B'_{0} = 1. \notag
\end{equation}

We can therefore write $B'_{n+1}$ as the sum:
\begin{equation}
    B'_{n+1}
    = \sum_{i=i_{min}}^{i_{max}}{\#\mathcal{P}_{i}([\![ 1,n+1 ]\!])}
    = \sum_{i=i_{min}}^{i_{max}} {\binom{n}{i} B'_{i}}
\notag
\end{equation}
and this establishes the recursive relation for $n \geqslant \sigma_{min}$. 

\vspace{1cm}

\noindent \textbf{Extended Stirling numbers. }
The Stirling number ${n \brace m}$ is the number of partitions of $n$ nodes in $m$ blocks. Its calculation follows the relations \shortcite{nielsen1906handbuch}:
\begin{gather}
    {0 \brace 0} = 1 \textnormal{, } {0 \brace n} = {n \brace 0} = 0 \textnormal{ for } n>0 \textnormal{, }
    {n+1 \brace m+1} = \sum_{i=m}^{n}{\binom{n}{i}{i \brace m}} \textnormal{ for } m>0.
\label{recurrence_stirling}
\end{gather}
Similarly, we can calculate $\psi_{\sigma_{min},\sigma_{max}}(n,m)$, the number of partitions in $m$ blocks when blocks sizes belong to $[\![\sigma_{min},\sigma_{max}]\!]$. First, there is no possible partition for $n<m\sigma_{min}$, therefore:
\begin{equation}
    \psi_{\sigma_{min},\sigma_{max}}(n,m) = 0 \textnormal{ for } 0<n<m\sigma_{min} 
\notag   
\end{equation}
The recursion then starts when there can be $m$ blocks of minimal size (i.e., $n=m\sigma_{min}$). To count possible partitions in this case, we first order all nodes in $n!$ different ways, and take each time the first group of $\sigma_{min}$ nodes, then the second group, and so on. Some partitions are counted several times as we have $m$ possible ways to order these groups and $\sigma_{min}!$ ways to order the nodes inside each group. The final count is:
\begin{equation}
    \psi_{\sigma_{min},\sigma_{max}}(n,m) = \frac{n!}{m!(\sigma_{min}!)^{m}} \textnormal{ for } n=m\sigma_{min} 
\notag 
\end{equation}

The terms in the sum of \eqref{recurrence_stirling} are the numbers of partitions where the node $(n+1)$ is in a group of size $(n+1-i)$ and the $i$ remaining nodes are partitioned in $m$ groups. As for $B'_{n}$ numbers, we can adapt the original recursive relation with the indexes \eqref{iminmax}:
\begin{equation}
    \psi_{\sigma_{min},\sigma_{max}}(n+1,m+1) = \sum_{i=i_{min}}^{i_{max}}{\binom{n}{i}\psi_{\sigma_{min},\sigma_{max}}(i,m)} \textnormal{ for } n \geqslant m\sigma_{min}.
\notag
\end{equation}

Finally, for the extreme case when the group of node $(n+1)$ is of size $(n+1)$ (i.e., $i=0$) and there are no left groups to form (i.e., $m=0$), we have to set for convenience:
\begin{equation}
    \psi_{\sigma_{min},\sigma_{max}}(0,0) = 1.
\notag   
\end{equation}

\newpage

\noindent\textbf{Appendix B. Translation of the Ewens distribution}\medskip

In this section, we demonstrate that the Ewens distribution \shortcite{ewens1972sampling}, as defined by Equation \eqref{equ:Ewens}, can be expressed in the form of the exponential family introduced in this paper with the following definition:
\begin{equation}
    \textnormal{Pr}_{\lambda}(P = p) = 
    \frac{\exp \Big( \log(\lambda) \: \#p + 
    \sum_{G \in p}{\log \big( (\#G-1)! \big) } \Big)}
    {\kappa_{\mathcal{P}} \big( (\log(\lambda),1) \big)}
\label{ewens3}
\end{equation}

To prove that this definition is equivalent to \eqref{equ:Ewens}, we develop its numerator and denominator. Following properties of the exponential and logarithm functions, the numerator can be expressed for any partition $P$:
\begin{equation}
    \exp(\log(\lambda)\#P + \sum_{G \in P}{\log((\#G-1)!)}
    = \lambda^{\#p} \; \prod_{G \in P}{\Gamma(\#G-1)!}
\label{conditions_ewens}
\end{equation}
Once this is established, proving the equivalence of the two definitions requires to prove that the normalizing constant of our model simplifies to:
\begin{equation}
    \kappa_{\mathcal{P}} \big( (\log(\lambda),1) \big)
    = \frac{\Gamma(n+\lambda-1)}{\Gamma(\lambda-1)}
\label{denominators_equality}
\end{equation}

Proving \eqref{denominators_equality} can be achieved by induction on $n$ the number of nodes. The distribution \eqref{ewens3} is defined for statistics that are functions of the group sizes, we can therefore define the sequence $\kappa_{n}$ relations found in Equations \eqref{recurrence_1} and \eqref{recurrence_2} of Appendix D.

From Equation \eqref{recurrence_1} and the property $\Gamma(1)=1$, we have for the basic case $n=1$:
\begin{equation}
    \kappa_{1} = \exp(\log(\lambda) + \log(1!) = \lambda 
\notag
\end{equation}
Besides, we know from properties of the Gamma function that $\Gamma(\lambda+1)=\lambda\Gamma(\lambda)$, we can therefore validate the relation \eqref{denominators_equality} for $n=1$:
\begin{equation}
    \kappa_{1} = \frac{\Gamma(\lambda)}{\Gamma(\lambda-1)}
\notag
\end{equation}

Let us now use the previously demonstrated formula \eqref{recurrence_2} for higher values of $n$. 
\begin{equation}
    \kappa_{n+1} 
    = \sum_{i=0}^{n}{ \binom{n}{i} \exp \Big( \log(\lambda) + \log\big( (n-i)! \big) \Big)  \kappa_{i} } 
    = \sum_{i=0}^{n}{ \binom{n}{i} \lambda (n-i)! \kappa_{i} } 
\label{def_kappa_n}
\end{equation}
We then separate this sum into two parts, one containing the term corresponding to $i=n$ and one containing the other terms:
\begin{equation}
    \kappa_{n+1} 
    = \lambda\kappa_{n} + \sum_{i=0}^{n-1}{ \binom{n}{i} \lambda(n-i)! \kappa_{i} } 
\notag
\end{equation}
Finally, we develop the binomial coefficients and re-arrange them in order to find the definition of $\kappa_{n}$ corresponding to the definition \eqref{def_kappa_n} for $(n+1)$:
\begin{align}
    \kappa_{n+1} 
    &= \lambda\kappa_{n} + \sum_{i=0}^{n-1}{ \frac{(n-1)!}{i!(n-1-i)!} \frac{n}{(n-i)} \lambda (n-i)! \kappa_{i} } \notag \\
    &= \lambda\kappa_{n} + n \bigg( \sum_{i=0}^{n-1}{ \binom{n-1}{i} \lambda \big( (n-1)-i \big)! \kappa_{i} } \bigg) \notag \\
    &= (\lambda + n)\kappa_{n} \notag
\end{align}

If we assume that \eqref{denominators_equality} holds for a given $n>0$, we therefore also have for $(n+1)$:
\begin{equation}
    \kappa_{n+1} 
    = (\lambda + n)\frac{\Gamma(n+\lambda-1)}{\Gamma(\lambda-1)} 
    = \frac{\Gamma((n+1)+\lambda-1)}{\Gamma(\lambda-1)} \notag
\end{equation}

This proves that the relation \eqref{denominators_equality} holds for any integer $n$ and that the model defined by \eqref{ewens3} is the same as the Ewens model defined by \eqref{equ:Ewens}.

\newpage

\noindent\textbf{Appendix C. Independence properties of the distribution}\bigskip

\noindent\textbf{Consistency.}
Let $P$ be a random partition over $\mathcal{A}$ following the distribution \eqref{equ:ERPM_jointform} and $P'$ a random partition over $\mathcal{A}'$ following the same distribution with identical sufficient statistics and parameters. We pose $\pi(P)$ the projection of $P$ over the subset $\mathcal{A}'$, and $\pi^{-1}(P')$ the set of partitions over the nodeset $\mathcal{A}$ whose projection is $P'$. 

Consistency of the distribution then implies equality between the marginal distribution of the random partition $P$ over $\mathcal{A}'$ and the distribution of $P'$. In other words, the family of projections of a partition model on $\mathcal{A}$ on the subset $\mathcal{A}'$ is a partition model with the same sufficient statistics and same parameters. This property translates to:
\begin{equation}
    \textnormal{Pr}_{\alpha} \big( P \in \pi^{-1}(p') \big) = \textnormal{Pr}_{\alpha} \big( P' = p' \big).  
\notag
\end{equation}

Here we present some counter-examples of distributions used in this paper for which this property does not hold. Let us use the space $\mathcal{A}=\{1,2,3\}$, its subset $\mathcal{A}'=\{1,2\}$, and the projection $\pi$ from $\mathcal{P}(\mathcal{A})$ to $\mathcal{P}(\mathcal{A}')$.

\medskip

\noindent \textit{Uniform model.}
Let us use the uniform distribution over $\mathcal{P}(\mathcal{A})$. There are $5$ different ways of partitioning this set, hence $1/5$ is the probability of any of these partitions. If we take a partition $p' = \big\{ \{1,2\} \big\}$, we can calculate the marginal probability:
\begin{equation}
    \textnormal{Pr} \Big( P \in \pi^{-1}( p') \Big)
    = \textnormal{Pr} \Big( P = \big\{ \{1,2,3\} \big\} \Big) +
    \textnormal{Pr} \Big( P = \big\{ \{1,2\},\{3\} \big\} \Big)
    = \frac{2}{5}
\notag    
\end{equation}
and the probability of observing $\big\{ \{1,2\} \big\}$ over $\mathcal{A}'$:
\begin{equation}
    \textnormal{Pr} \Big( P' = p' \Big) =
    \textnormal{Pr} \Big( P' = \big\{ \{1,2\} \big\} \Big) = \frac{1}{2}
\notag
\end{equation}
The uniform distribution is therefore not consistent.

\medskip

\noindent \textit{Model with one statistic $s_1(P)=\#P$.}  We can use again the same example on the same sets and $p' = \big\{ \{1,2\} \big\}$. We have as marginal probability:
\begin{gather}
    \textnormal{Pr}_{\alpha_1} \Big( P \in \pi^{-1}(p') \Big) 
    = \textnormal{Pr}_{\alpha_1} \Big( P = \big\{ \{1,2,3\} \big\} \Big) +
    \textnormal{Pr}_{\alpha_1} \Big( P = \big\{ \{1,2\},\{3\} \big\} \Big) \notag \\
    = \frac{\exp(\alpha_1) + \exp(2\alpha_1)}{\exp(\alpha_1) + 3\exp(2\alpha_1)+\exp(3\alpha_1)}
\notag    
\end{gather}
and:
\begin{equation}
    \textnormal{Pr}_{\alpha_1} \Big( P' = p' \Big) =
    \textnormal{Pr}_{\alpha_1} \Big( P' = \big\{ \{1,2\} \big\} \Big) = \frac{\exp(\alpha_1)}{\exp(\alpha_1)+\exp(2\alpha_1)}
\notag
\end{equation}
Having these two terms equal is equivalent to the equation $\exp(2\alpha_1) = 0$, which has no solution in $\mathbb{R}$. Again the consistency condition cannot be fulfilled for such models.

\medskip

\noindent\textbf{Neutrality.}
We show in this section that the neutrality property defined by Equation \eqref{neutrality} holds for any model defined for a set of statistics of the form:
\begin{equation}
    s_{k}(P) = \sum_{G \in P}{f_{k}(G)}
\notag
\label{functions_of_groups}
\end{equation}
with $(f_{k})$ defined as real functions of the groups in the partition (i.e. representing any characteristic of the group). This definition covers all statistics used in this article, however, other types of statistics could also lead to neutral distributions.

Let $P$ be a random partition defined for such a model with parameter vector $\alpha$. Furthermore, let $p$ be the observed partition that has the property of being the union of its projections over the subsets $\mathcal{A}$ and $\mathcal{A}^{c}$. We can write:
\begin{gather}
      \textnormal{Pr}_{\alpha} \Big( P = p \: | \: P = \pi(P) \cup \pi^{c}(P)\Big) = 
     \frac
     {\textnormal{Pr}_{\alpha} \Big( P = p \textnormal{, } p = \pi(p) \cup \pi^{c}(p) \Big) }
     {\textnormal{Pr}_{\alpha} \Big( P = \pi(P) \cup \pi^{c}(P) \Big) }
\notag
\end{gather}

Since the observed partition verifies $p = \pi(p) \cup \pi^{c}(p)$, the numerator simplifies to:
\begin{equation}
   \textnormal{Pr}_{\alpha} \Big( P = p \textnormal{, } p = \pi(p) \cup \pi^{c}(p) \Big)
   = \textnormal{Pr}_{\alpha} \Big( P = p \Big)
   \notag
\end{equation}
and since summing over all groups of $p$ is equivalent to summing over the groups in $\pi(p)$ and $\pi^{c}(p)$, this probability factorizes as follows:
\begin{gather}
   \textnormal{Pr}_{\alpha} \Big( P = p \Big) 
   = \frac{1}{\kappa_{\mathcal{P}(\mathcal{A})}(\alpha)} \exp \bigg( \sum_{k \in K}{\alpha_{k} 
   \Big( \sum_{G \in \pi(p)} {f_{k}(G)} + \sum_{G \in \pi^{c}(p)} {f_{k}(G)} \Big) \bigg)} \notag \\
   = \frac{1}{\kappa_{\mathcal{P}(\mathcal{A})}(\alpha)} \exp \bigg( \sum_{k \in K}{\alpha_{k} 
   \sum_{G \in \pi(p)} {f_{k}(G)} \bigg)}
   \exp \bigg( \sum_{k \in K}{\alpha_{k} 
   \sum_{G \in \pi^{c}(p)} {f_{k}(G)} \bigg)}. 
\label{neutrality_num}
\end{gather}

The denominator expresses the probability of having the random partition $P$ verifying $P = \pi(P) \cup \pi^{c}(P)$. It is the sum of probabilities of all partitions in $\mathcal{P}(\mathcal{A})$ with this property. If we define $\mathcal{Q}(\mathcal{A},\mathcal{A}')$ the set of these partitions, we can define a bijection $b: \mathcal{Q}(\mathcal{A},\mathcal{A}') \rightarrow \big( \mathcal{P}(\mathcal{A}'), \mathcal{P}(\mathcal{A}'^{c})\big)$ such that $b(P) = (\pi_{\mathcal{A}'}(P),\pi_{\mathcal{A}'^{c}}(P))$. We deduce:
\begin{gather}
    \textnormal{Pr}_{\alpha} \Big( P = \pi(P) \cup \pi^{c}(P) \Big) \notag \\
    = \sum_{\widetilde{P} \in \mathcal{Q}(\mathcal{A},\mathcal{A}')}
    {\frac{1}{\kappa_{\mathcal{P}(\mathcal{A})}(\alpha)}
    \exp \bigg( \sum_{k \in K}{\alpha_{k} 
   \Big(\sum_{G \in \pi_{\mathcal{A}'}(\widetilde{P})} {f_{k}(G)} +\sum_{G \in \pi_{\mathcal{A}'^{c}}(\widetilde{P})} {f_{k}(G)} \Big) } \bigg)} \notag \\
   = \sum_{\widetilde{P}_1 \in \mathcal{P}(\mathcal{A}')} \sum_{\widetilde{P}_2 \in \mathcal{P}(\mathcal{A}'^{c})}
    {\frac{1}{\kappa_{\mathcal{P}(\mathcal{A})}(\alpha)}
    \exp \bigg( \sum_{k \in K}{\alpha_{k} 
   \Big(\sum_{G \in \widetilde{P}_1} {f_{k}(G)} +\sum_{G \in \widetilde{P}_2} {f_{k}(G)} \Big) } \bigg)} \notag \\
   = \frac{1}{\kappa_{\mathcal{P}(\mathcal{A})}(\alpha)}
   \bigg( \sum_{\widetilde{P}_1 \in \mathcal{P}(\mathcal{A}')} {\exp \bigg( \sum_{k \in K}{\alpha_{k} 
   \sum_{G \in \widetilde{P}_1} {f_{k}(G)} } \bigg)}
   \bigg)
   \bigg( \sum_{\widetilde{P}_2 \in \mathcal{P}(\mathcal{A}'^{c})} {\exp \bigg( \sum_{k \in K}{\alpha_{k} 
   \sum_{G \in \widetilde{P}_2} {f_{k}(G)} } \bigg)}
   \bigg) \notag 
\end{gather}
and we can simplify:
\begin{equation}
    \textnormal{Pr}_{\alpha} \Big(P = \pi(P) \cup \pi^{c}(P) \Big)
   = \frac{\kappa_{\mathcal{P}(\mathcal{A}')}(\alpha)
   \kappa_{\mathcal{P}(\mathcal{A}'^{c})}(\alpha)}
   {\kappa_{\mathcal{P}(\mathcal{A})}(\alpha)}.
\label{neutrality_den}
\end{equation}

By dividing the terms \eqref{neutrality_num} and \eqref{neutrality_den}, the term $\kappa_{\mathcal{P}(\mathcal{A})}$ simplifies and we finally have:
\begin{gather}
    \textnormal{Pr}_{\alpha} \Big(  P = p \: | \: P = \pi(P) \cup \pi^{c}(P) \Big) \notag \\
    = \frac{1}{\kappa_{\mathcal{P}(\mathcal{A}')}(\alpha)}
    \exp \bigg( \sum_{k \in K}{\alpha_{k} 
   \sum_{G \in \pi(p)} {f_{k}(G)} \bigg)}
    \times 
    \frac{1}{\kappa_{\mathcal{P}(\mathcal{A}'^{c})}(\alpha)}
    \exp \bigg( \sum_{k \in K}{\alpha_{k} 
   \sum_{G \in \pi^{c}(p)} {f_{k}(G)} \bigg)} \notag \\
    = \textnormal{Pr}_{\alpha} \Big( \pi(P) = \pi(p) \Big)
    \times \textnormal{Pr}_{\alpha} \Big( \pi^{c}(P) = \pi^{c}(p) \Big) \notag
\end{gather}
and this demonstrates the property of neutrality as defined by \eqref{neutrality}.

\newpage

\noindent\textbf{Appendix D. Calculation of the normalizing constant when statistics are functions of block sizes}

\bigskip

Here, we demonstrate that the normalizing constant $\kappa$ as expressed by Equation \eqref{equ:ERPM_kappa} can be calculated with a recursive formula when sufficient statistics $\big(s_{k}\big)_{k\in K}$ are of the form:
\begin{equation}
    s_{k}(P) = \sum_{G \in P}{f_{k}(\#G)}
\label{functions_of_sizes}
\end{equation}
with $(f_{k})$ defined as functions from the set of possible group sizes to $\mathbb{R}$. In the rest of the proof, we also pose:
\begin{equation}
    f(P) = \exp \Big( \sum_{k \in K}\alpha_{k}\sum_{G \in P}{f_{k}(\#G)} \Big).
\label{functions_of_sizes_gen}
\end{equation}

In such cases, the probability distribution defined by \eqref{equ:ERPM_jointform} is said to be exchangeable \shortcite{mccullagh2011random}, as it is invariant under any permutation of the nodes. The normalizing constant $\kappa$ then only depends on $n$ and is noted $\kappa_n$.

For the sake of conciseness, we demonstrate the relation \eqref{equ:kappa_recurrence_restricted} for the constant $\kappa'_{n}$ defined over the set $\mathcal{P}'([\![ 1,n ]\!])$ with groups sizes between $\sigma_{min}$ and $\sigma_{max}$. The relation \eqref{kappa_recurrence} for the general case directly follows.

The proof is based on a similar logic to the one used in Appendix A. To initialize the recursion, we know that there are no possible partitions for smaller sizes, and that there is only one partition with one group for $n=\sigma_{min}$. Therefore:
\begin{gather}
    \kappa'_{n} = 0 \textnormal{ for } n < \sigma_{min} \notag \\
    \kappa'_{n} = \exp \Big( \sum_{k \in K}{\alpha_{k}f_{k}(\sigma_{min})} \Big) \textnormal{ for } n = \sigma_{min} 
\label{recurrence_1}
\end{gather}

For $n>\sigma_{min}$, we can use the subsets $\mathcal{P}_{i}([\![ 1,n+1 ]\!])$ defined by \eqref{subsetsi} and write:
\begin{equation}
    \kappa'_{n+1} = \sum_{\widetilde{P} \in \mathcal{P}'([\![ 1,n+1 ]\!])}{ f(\widetilde{P}) }
    = \sum_{i=i_{min}}^{i_{max}}{\Bigg( \sum_{\widetilde{P} \in \mathcal{P}_{i}([\![ 1,n+1 ]\!])}{  f(\widetilde{P}) }} \Bigg).
\notag
\end{equation}

Let us define $\mathcal{G}_{i}([\![ 1,n+1 ]\!])$ the set of all possible groups of nodes in $[\![ 1,n+1 ]\!]$ that include the node $(n+1)$ and whose size is equal to $(n+1-i)$. To enumerate all possible partitions of $\mathcal{P}_{i}([\![ 1,n+1 ]\!])$, we enumerate all groups in $\mathcal{G}_{i}([\![ 1,n+1 ]\!])$ and all possible partitions over the remaining $i$ nodes. With this notation, we have:
\begin{equation}
    \kappa'_{n+1} 
    = \sum_{i=i_{min}}^{i_{max}}{\Bigg( \sum_{g \in \mathcal{G}_{i}([\![ 1,n+1 ]\!])}{\sum_{\widetilde{P} \in \mathcal{P}'([\![ 1,n+1 ]\!] \setminus g)}{ f(\widetilde{P} \cup g) }} } \Bigg).
\notag
\end{equation}

Since the definition of the function $f$ is invariant under permutations of the nodes, we can re-order the $i$ remaining nodes from $1$ to $i$. From this we deduce that for any group $g \in \mathcal{G}_{i}([\![ 1,n+1 ]\!])$ there is a bijection $b_{g}:\mathcal{P}'([\![ 1,n+1 ]\!] \setminus g) \rightarrow \mathcal{P}'([\![ 1,i ]\!])$ such that partitions over remaining nodes are defined for these re-ordered nodes. We can therefore replace the sum indices in the previous expression: 
\begin{equation}
    \kappa'_{n+1} 
    = \sum_{i=i_{min}}^{i_{max}}{\Bigg( \sum_{g \in \mathcal{G}_{i}([\![ 1,n+1 ]\!])}{\sum_{\widetilde{P} \in \mathcal{P}'([\![ 1,i ]\!])}{ f(\widetilde{P} \cup g) }} } \Bigg).
\notag
\end{equation}

We can the use the definition \eqref{functions_of_sizes} of the statistics $s_k$ to derive:
\begin{equation}
    f(\widetilde{P} \cup g) = \exp \bigg( \sum_{k \in K}{\alpha_{k} \Big( \sum_{G \in \widetilde{P}}{f_{k}(\#G)} } \bigg) \exp \bigg( \sum_{k \in K}{\alpha_{k} f_{k}(\#g) \Big) } \bigg) = f(\widetilde{P})f(g)
\notag
\end{equation}
and factorize:
\begin{equation}
    \kappa'_{n+1} 
   = \sum_{i=i_{min}}^{i_{max}}{\Bigg( \sum_{g \in \mathcal{G}_{i}([\![ 1,n+1 ]\!])}{ f(g) \sum_{\widetilde{P} \in \mathcal{P}'([\![ 1,i ]\!])}{ f(\widetilde{P})  }}  \Bigg)}.
\notag 
\end{equation}

By definition, the following term simplifies to one of the previously calculated normalizing constants:
\begin{equation}
    \sum_{\widetilde{P} \in \mathcal{P}'([\![ 1,i ]\!])}{ f(\widetilde{P}) } = \kappa'_{i}, 
\notag
\end{equation}
except in the case of $i=0$ for which we set:
\begin{equation}
    \kappa'_{0} = 1.
\end{equation}

Moreover, we know that for any $g \in \mathcal{G}_{i}([\![ 1,n+1 ]\!])$, $f_{k}(\#g)=f_{k}(n+1-i)$. Developing $f(g)$ then removes any term depending on $g$. The size of $\mathcal{G}_{i}([\![ 1,n+1 ]\!])$ being the number of ways to choose $n-i$ elements (or $i$ elements) among $n$ nodes, we deduce:
\begin{equation}
    \kappa'_{n+1} 
    = \sum_{i=i_{min}}^{i_{max}}{ \binom{n}{i} \exp \bigg( \sum_{k \in K}{\alpha_{k} f_{k}(n+1-i)} \bigg)  \kappa'_{i} }.  
\label{recurrence_2}
\end{equation}

These expressions show that we can recursively construct the sequence $\kappa'_n$, using the initialization \eqref{recurrence_1} and the recursive relation \eqref{recurrence_2}. Given that the relation is linear and its factors are easy to calculate for a reasonable number of nodes, these normalizing constants can be directly calculated.

\appendix

\end{document}